%% file: processing_networks_estimation_RL.tex
\documentclass[twoside]{IEEEtran}
\usepackage[english]{babel}
\usepackage[utf8]{inputenc}
\usepackage[T1]{fontenc}

\usepackage{microtype}
\usepackage{setspace}
\usepackage{blindtext}
\usepackage{siunitx}
\usepackage{comment}
\usepackage{makecell}
\usepackage{xstring}
\usepackage{orcidlink}

\usepackage{titlecaps}
\Addlcwords{or the with if a an of for and to -off off in not by via}
\usepackage[caption=false]{subfig}
\usepackage{graphicx}
\usepackage{fancyhdr} 
\usepackage{color}
\usepackage{epsfig}
\graphicspath{{Images/}}
\usepackage{tikz}
\usepackage{pgfplots,pgfplotstable}
\usetikzlibrary{patterns,fit,matrix}
\usepackage{tikzscale}
\usetikzlibrary{positioning}
\usepackage{scalerel}
\usetikzlibrary{arrows,
	patterns,
	plotmarks,
	svg.path,
	shapes.multipart}
\usepgfplotslibrary{fillbetween}
\pgfplotsset{compat=newest,
	compat/bar nodes=1.8,
	every axis/.append style={
		label style={font=\Large},
		tick label style={font=\large} 
	}
}
\tikzstyle{int}=[draw, fill=black!10, minimum size=5em,thick]
\tikzstyle{init} = [pin edge={to-,thick,black}]
\usepackage{booktabs}
\usepackage{multirow}

\usepackage{array}
\usepackage{enumitem}
\usepackage[bottom]{footmisc}

\usepackage{amsthm,amsmath,amssymb,amsfonts}
\usepackage{relsize}
\usepackage{nicefrac}
\usepackage{bbm}
\usepackage{mathtools}
\usepackage[mathscr]{eucal}
\usepackage[short,c2]{optidef}

\usepackage{url}
\usepackage{hyperref}
\usepackage[capitalize]{cleveref}

\usepackage{cite}

\input{mycommands}

\title{\titlecap{
	to compute or not to compute?
	adaptive smart sensing
	in resource-constrained edge computing}}

\author{Luca~Ballotta\,\textsuperscript{\orcidlink{0000-0002-6521-7142}},~\IEEEmembership{Member,~IEEE}, %
		Giovanni~Peserico\,\textsuperscript{\orcidlink{0000-0001-5444-6946}}, %
		Francesco~Zanini\,\textsuperscript{\orcidlink{0000-0003-3036-0675}}, and %
		Paolo~Dini\,\textsuperscript{\orcidlink{0000-0001-6756-0289}} %
	\thanks{This work was supported in part by the Italian Ministry of Education, University and Research (MIUR) 
		through the PRIN project under Grant 2017NS9FEY titled ``Realtime Control of 5G Wireless Networks''
		and through the initiative ``Departments of Excellence'' (Law 232/2016),
		in part by the Spanish Project PID2020-113832RB-C22(ORIGIN)/MCIN/AEI/10.13039/50110001103,
		and in part by the Ministerio de Asuntos Econ\'omicos y Transformaci\'on Digital and the European Union-NextGenerationEU 
		in the frameworks of the Plan de Recuperaci\'on, Transformaci\'on y Resiliencia and of the Mecanismo de Recuperaci\'on y Resiliencia under references TSI-063000-2021-18/24/77 (6GOASIS).
		The views and opinions expressed in this work are those of the authors and do not necessarily reflect those of the funding institutions.}%
	\thanks{Luca Ballotta and Giovanni Peserico
		are with the Department of Information Engineering, University of Padova, 35131 Padova, Italy
		(e-mail: ballotta@dei.unipd.it; giovanni.peserico@phd.unipd.it) .}
	\thanks{Francesco Zanini is with the Department of Computing Science, University of Alberta, T6G 2E8 Edmonton, Alberta, Canada
		(e-mail: fzanini@ualberta.ca).}%
	\thanks{Paolo Dini is with the Centre Tecnologic de Telecomunicacions de Catalunya, 08860 Barcelona, Spain
		(e-mail: paolo.dini@cttc.es).}%
	\thanks{The first three authors contributed equally.}
}

\begin{document}
	
	\bstctlcite{bib-options}
	\maketitle
	\input{Sections/abstract}
	\input{Sections/introduction}
			\input{Sections/related-work}
			\input{Sections/contribution}
	\input{Sections/setup}
			\input{Sections/system-model}
					\input{Sections/smart-sensors}

					\input{Sections/wireless-channel}
					\input{Sections/base-station}
			\input{Sections/problem-formulation}
	\input{Sections/centralized-implementation}
			\input{Sections/homogeneous-sensors}
			\input{Sections/heterogeneous-sensors}
	\input{Sections/RL-formulation}
			\input{Sections/RL-tradeoff}
			\input{Sections/RL-discussion}
	\input{Sections/simulations}
	        \input{Sections/sim-python}
			\input{Sections/sim-omnet}
			\input{Sections/sim-discussion}
	\input{Sections/conclusion}
	
	
	\input{processing_networks_estimation_RL.bbl}

	\input{Bio/bios}

	\clearpage
	\newpage
	\appendices
	\input{Appendix/kalman-filter}
	\input{Appendix/drone-sim-extra}
	\clearpage
	\newpage
    \input{Appendix/Qconvergence}

\end{document}

%% file: mycommands.tex
\newcommand{\eg}{\emph{e.g.,}\xspace}
\newcommand{\ie}{\emph{i.e.,}\xspace}

\newcommand{\cf}{cf.\xspace}

\newcommand{\tradeoff}{latency-accuracy trade-off\xspace}
\newcommand{\homProcPol}{homogeneous sensing policy\xspace}
\newcommand{\netProcPol}{network sensing policy\xspace}
\newcommand{\omnet}{OMNeT++}
\newcommand{\raw}{\mathrm{r}}
\newcommand{\proc}{\mathrm{p}}
\newcommand{\sleep}{\mathrm{s}}

\newcommand{\Real}[1]{ { {\mathbb R}^{#1} } }

\newcommand{\inv}{^{-1}}

\newcommand{\tr}[1]{\mathrm{Tr}\left(#1\right)}

\newcommand{\var}[1]{\mathrm{Var}\left(#1\right)}

\newcommand{\gauss}{\mathcal{N}}

\newcommand{\x}[1]{x_{#1}}
\newcommand{\w}[1]{w_{#1}}

\newcommand{\yi}[2]{y_{#1}^{(#2)}}
\newcommand{\vi}[2]{v_{#1}^{(#2)}}
\newcommand{\V}[1]{V_{#1}}
\newcommand{\delayRaw}[1][\@empty]{\tau_%
	\ifx\@empty#1 {\text{raw}} \else {{#1,\text{raw}}} \fi}
\newcommand{\delayProc}[1][\@empty]{\tau_%
	\ifx\@empty#1 {\text{proc}} \else {{#1,\text{proc}}} \fi}
\newcommand{\delayComm}[1][]{\delta_{#1}}
\newcommand{\delayCommRaw}[1][\@empty]{\delta_%
	\ifx\@empty#1 {\text{raw}} \else {{#1,\text{raw}}} \fi}
\newcommand{\delayCommProc}[1][\@empty]{\delta_%
	\ifx\@empty#1 {\text{proc}} \else {{#1,\text{proc}}} \fi}
\newcommand{\delayRec}[1]{\Delta_{#1}}
\newcommand{\delayRecRaw}[1][\@empty]{\Delta_%
	\ifx\@empty#1 {\text{raw}} \else {{#1,\text{raw}}} \fi}
\newcommand{\delayRecProc}[1][\@empty]{\Delta_%
	\ifx\@empty#1 {\text{proc}} \else {{#1,\text{proc}}} \fi}

\newcommand{\varRaw}[1][\@empty]{V_%
	\ifx\@empty#1 {\text{raw}} \else {{#1,\text{raw}}} \fi}
\newcommand{\varRawS}[1][\@empty]{v_%
	\ifx\@empty#1 {\text{raw}} \else {{#1,\text{raw}}} \fi}
\newcommand{\varProc}[1][\@empty]{V_%
	\ifx\@empty#1 {\text{proc}} \else {{#1,\text{proc}}} \fi}
\newcommand{\varProcS}[1][\@empty]{v_%
	\ifx\@empty#1 {\text{proc}} \else {{#1,\text{proc}}} \fi}
\newcommand{\samplingSequence}[1]{\mathcal{K}_{#1}}
\newcommand{\measurements}[2][]{\mathcal{Y}_{#2}^{#1}}
\newcommand{\allmeasurements}[1]{\mathcal{Y}_{#1}}
\newcommand{\delayFus}[1]{\phi_{#1}}
\newcommand{\decision}[2]{\gamma_{#1}^{#2}}
\newcommand{\decisionDelay}[1]{\tau^{(#1)}}
\newcommand{\decisionTimestamps}{\mathcal{K}}

\newcommand{\kell}[2][]{k^{(\ell#1)}_{#2}}
\newcommand{\kL}[1][]{k^{(L)}_{#1}}
\newcommand{\policy}[1][]{\pi_{#1}}
\newcommand{\policyHom}[1][\@empty]{\pi_%
	\ifx\@empty#1 {\text{hom}} \else {{\text{hom},#1}} \fi}
\newcommand{\decisionHom}[2][\@empty]{\gamma_{#2}^%
	\ifx\@empty#1 \text{hom} \else {\text{hom},#1} \fi}

\newcommand{\np}{n_\text{p}}
\newcommand{\ns}{n_\text{s}}
\newcommand{\policyNet}{\pi_\text{net}}

\newcommand{\sensSet}{\mathcal{V}}

\newcommand{\xhat}[2]{\hat{x}_{#1}^{#2}}
\newcommand{\xtilde}[2]{\tilde{x}_{#1}^{#2}}
\newcommand{\Pmat}[2]{P_{#1}^{#2}}
\newcommand{\pred}[2]{\mathcal{P}^{#2}\left(#1\right)}
\newcommand{\update}[2]{\mathcal{U}\left(#1,#2\right)}

\newcommand{\kalman}[1]{f_{\text{Kalman}}\left(#1\right)}

\newcommand{\epsmax}{\epsilon_0}
\newcommand{\epsmin}{\epsilon_{\text{min}}}
\newcommand{\discFact}{\gamma}
\newcommand{\learnRate}{\alpha}

\theoremstyle{plain}

\theoremstyle{definition}
\newtheorem{definition}{Definition}
\newtheorem{prob}{Problem}
\newtheorem{ass}{Assumption}
\theoremstyle{remark}
\newtheorem{rem}{Remark}

\newcommand{\lr}{\left(}
\newcommand{\rr}{\right)}
\newcommand{\ls}{\left[}
\newcommand{\rs}{\right]}
\newcommand{\lb}{\left\lbrace}
\newcommand{\rb}{\right\rbrace}

\addto\captionsenglish{\renewcommand{\figurename}{Fig.}}
\addto\captionsenglish{\renewcommand{\tablename}{TABLE}}

\newcommand{\LB}[1]{{\color{blue}LB: #1}} 
\newcommand{\blue}[1]{{\color{blue}#1}}
\newcommand{\red}[1]{{\color{red}#1}}
\newcommand{\revision}[2][]{#2\xspace}

\newcommand{\linkToPdf}[1]{\href{#1}{\blue{(pdf)}}}
\newcommand{\linkToPpt}[1]{\href{#1}{\blue{(ppt)}}}
\newcommand{\linkToCode}[1]{\href{#1}{\blue{(code)}}}
\newcommand{\linkToWeb}[1]{\href{#1}{\blue{(web)}}}
\newcommand{\linkToVideo}[1]{\href{#1}{\blue{(video)}}}
\newcommand{\linkToMedia}[1]{\href{#1}{\blue{(media)}}}
\newcommand{\award}[1]{\xspace} 
\addto\extrasenglish{}
\addto\extrasenglish{}
\addto\extrasenglish{}

%% file: Sections/abstract.tex

\begin{abstract}
	We consider a network of smart sensors for an edge computing application
	that sample a \revision{time-varying} signal
	and send updates to a base station for remote global monitoring. 
	Sensors are equipped with sensing and compute,
	and can either send raw data or process them on-board before transmission.
	Limited hardware resources at the edge generate a fundamental \emph{\tradeoff}:
	raw measurements are inaccurate but timely,
	whereas \emph{accurate}
	processed updates are available after \emph{\revision{processing} delay}.
	Hence,
	one needs to decide when sensors should transmit raw measurements 
	or rely on local processing to maximize network \revision{monitoring} performance.
	To tackle this sensing design problem,
	we model an estimation-theoretic optimization framework that embeds \revision{both} computation and communication latency,
	and propose a Reinforcement Learning-based approach
	that dynamically allocates computational resources at each sensor.
	Effectiveness of our proposed approach is validated through numerical \revision{experiments
	motivated by smart sensing for} the Internet of Drones and self-driving vehicles.
	\revision{In particular,
	we show that,
	under constrained computation at the base station,
	monitoring performance can be further improved by an \textit{online sensor selection}.}
	
	\begin{IEEEkeywords}
		Communication latency, computation latency, edge computing, Q-learning, resource allocation, sensing design.
	\end{IEEEkeywords}
\end{abstract}

%% file: Sections/introduction.tex

\section{Introduction}

\IEEEPARstart{D}{istributed} computation scenarios such as
the Internet of Things and Industry 4.0 
represent a major breakthrough in engineering applications,
whereby coordination of sensing and actuation moves away from classical centralized controllers
to servers and devices at the network edge.
This empowers multiple local systems to achieve together complex goals at global level:
this happens with management of electricity and energy harvesting in smart grids~\cite{8636257,ERDEM201898},
resource utilization in smart agriculture~\cite{9316211,s20072081},
modularization and productivity enhancement in Industry 4.0~\cite{BUENO2020106774,IVANOV2018134,KRUGH201889}, 
urban traffic with interconnected vehicles~\cite{doi:10.2514/6.2020-0602,9072289},
\revision[2]{and space-air-ground services~\cite{9748064}}.

In particular,
recent advances in both embedded electronics,
with powerful micro controllers and GPU processors~\cite{9613590,9156208},
and new-generation communication protocols for massive networks,
such as 5G~\cite{li20185g,9634111},
are currently pushing network systems to rely on 
sensors and,
more in general, 
edge devices to carry most of the computational burden. 
Indeed,
distributed computation paradigms such as edge and fog computing~\cite{yi2015survey,shi2016edge,ogrady_edge_2019,chen_internet_2019} \revision[2]{\cite{9679802}} 
and federated and decentralized learning~\cite{9154332,9337227,9562559},
even though still in their infancy,
enjoy febrile activity and excitement across the research community.

Despite the growing resources and technological development,
emerging edge technologies are still limited compared to centralized servers:
indeed,
edge devices are forced to trade several factors,
such as hardware cost for processing speed for energy consumption.
In particular,
data processing on devices at the edge requires a non-negligible computational time.

In this work,
we consider a group of edge smart sensors,
such as compute-equipped IoT nodes or UAVs,
that measure a signal of interest 
-- \eg voltage in a smart grid,
or movements of vehicles for surveillance --
and transmit
\revision{the measurements}
to a base station that performs remote global monitoring and possibly decision-making.
Limited hardware resources induce a \emph{\tradeoff} at each sensor, 
that can supply either raw,
inaccurate samples of the monitored signal 
or 
\revision{refine those same data on-board by
	running suitable algorithms,
	which produce high-quality measurements}
at the cost of \emph{processing delay} caused by constrained \revision{hardware}.
\revision{Such \textit{local processing}}
may consist of averaging or filtering \revision{a batch} of noisy samples, 
or \revision{feature extraction}
from images or other high-dimensional data~\cite{fang_fpga-based_2017,9561090}, 
to mention a few examples.
Because the monitored system evolves dynamically,
delays in transmitted measurements may hinder usefulness of these in real-time tasks, 
so that sensing design for multiple, 
heterogeneous sensors
becomes challenging. 
In particular, as sensors cooperate,
it is unclear which of them should rely on local computation to transmit accurate information,
and which ones would be better off sending raw data. 
Also, channel constraints such as limited bandwidth \revision{may} introduce non-negligible \emph{communication latency},
further increasing complexity of the sensing design. 
Specifically,
local processing might compress acquired samples,
so that transmission of raw data to the base station takes longer.

%% file: Sections/related-work.tex

\subsection{Related Literature}
Resource allocation in terms of sensor and actuator selection represents a major research topic in IT, robotics, and control.
Classically,
the need for selection emerges from maximization of a performance metric subject to limited resource budget,
being it of economical, 
functional (\eg weight of autonomous platforms), 
spatial (\eg locations to place sensors),
or other nature.
Typical works in this field%
~\cite{anvaripour_novel_2020,
	mao_effectiveness_2018,
	brunello_virtual_2021,
	devos_sensor_2021,
	schon_integrity_2018,
	li_integrating_2008,
	clark_multi-fidelity_2021,
	alonso_smart_2020,
	khan_adaptive_2021}
focus on such budget-related constraints and pay little attention to impact on system dynamics.
For example,~\cite{anvaripour_novel_2020} proposes selection strategies based on coverage probability and energy consumption for a target tracking problem,
\cite{khan_adaptive_2021} studies a clustering-based selection to address communication constraints in underwater environments,
and~\cite{clark_multi-fidelity_2021} tackles placement of cheap and expensive sensors to optimize reconstruction of dynamical variables.
The aforementioned works, even though address computation and/or communication issues,
either care about energy consumption or address latency in a qualitative way,
but do not use that information to compute an exact performance metric that depends on the system dynamics.
Another, more control theoretic, body of work
exploits tools from set-valued optimization, \eg submodular functions with matroid constraints~\cite{jawaid_submodularity_2015},
or studies analytical bounds~\cite{gupta_stochastic_2006} or convex formulations~\cite{maity_sensor_2022,li_integrating_2008},
yet within a static framework that does not address changes in the overall dynamics.

In a similar realm,
control theory is traditionally concerned with either channel-aware estimation and control,
or co-design of communication and controller,
addressing wireless channel issues such as unreliability, latency, and more in general limited information%
~\cite{schenato_foundations_2007,chiuso_lqg_2013,tatikonda_control_2004,le_ny_differentially_2014,shafieepoorfard_rational_2013,nair_stabilizability_2004}.
For example,~\cite{schenato_foundations_2007,nair_stabilizability_2004} are concerned with rate-constrained stabilizability,
while~\cite{chiuso_lqg_2013,shafieepoorfard_rational_2013} address LQR and LQG control.
More recently, performance of wireless cyber-physical systems subject to state and input constraints
has been thoroughly investigated leveraging model-based prediction and  optimization tools such as MPC%
~\cite{pezzutto_adaptive_2020,park_wireless_2018,branz_drive-by-wi-fi_2022,li_network-based_2014}.
However, also this line of work does not consider processing-dependent delays and their effect on dynamics and performance.
Even in recent work on sensing, 
communication,
and control co-design~\cite{9099596,9654960}
there is no unifying framework that exactly relates sensing and computation on resource-limited platforms to
estimation and control performance in dynamical networks. 
\revision{A novel framework concerned with an adaptive design for LQG control
	which addresses accuracy-dependent sensing latency
	is presented in~\cite{lopez2020cdc_latencyScheduling}.
	However,
	it considers a single sensor and proposes a heuristic solution with no theoretical guarantees.
}

A recent body of literature tailored to edge and fog computing studies distributed computation on resource-constrained devices,
focusing on minimization of 
delays~\cite{ruthDemoAbstractInnetwork2018,
	linTimeDrivenDataPlacement2019,
	Taami19wfcs-experimentalLatency,
	kaoHermesLatencyOptimal2017}
or latency-dependent energy consumption~\cite{lyuEnergyEfficientAdmissionDelaySensitive2018}.
While there is a clear, empirically supported intuition that outdated sensory information
is detrimental to performance through the dynamical nature of monitored systems,
the above works do not address the true performance metrics
(which may be unknown or too complicated to compute),
but employ \revision{heuristic proxies (\eg delays)
without quantifying impact on closed-loop performance.

Finally, 
a similar trend is found within a recent body of the communications literature on
Age of Information (AoI)~\cite{Yates21jsac-aoi,
	Kosta20tc-nonlinearAoI,
	Tripathi21tmc-aoiGraphs,
	Ornee21tn-aoi},
a metric that quantifies the time elapsed since the latest received update
from a source of information.
These works focus on minimizing quantities related to AoI of updates,
but typically neglect dynamics of the measured variables.
Also,
most works,
\eg~\cite{9589966,Kadota21tmc-aoiWirelessNetwork},
assume that the dynamical systems measured by different sensors are uncoupled,
limiting applicability of this approach in networked control systems.}

%% file: Sections/contribution.tex

\subsection{Novel Contribution and Organization of the Article}

\revision{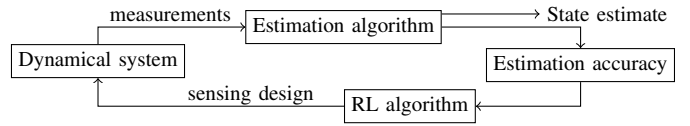
\begin{figure}
		\centering
		\begin{tikzpicture}
			\begin{scope}[shift={(0,0)},scale=1.1]
				\footnotesize			
				\node [draw] (sys) at (1,0) {Dynamical system};
				\node [above right=.2cm and -1.1cm of sys] (meas) {measurements};
				\node [draw, above right=-.0cm and .8cm of sys] (kf) {Estimation algorithm};
				\draw [->, to path={|- (\tikztotarget)}] (sys) edge (kf);
				\node [above right= -.25cm and 1.3cm of kf] (est) {State estimate};
				\draw [<-] (est) edge (est -| kf.east);
				\node [draw, below right=-.4cm and 4cm of sys] (cov) {Estimation accuracy};
				\draw [->, to path={-| (\tikztotarget)}] (kf) edge (cov);
				\node [draw, below right=.6cm and -1.3cm of kf] (rl) {RL algorithm};
				\draw [->, to path={|- (\tikztotarget)}] (cov) edge (rl);
				\draw [->, to path={-| (\tikztotarget)}] (rl) edge (sys);
				\node [above left=-.3cm and .3cm of rl] (sens) {sensing design};
			\end{scope}
		\end{tikzpicture}
		\caption{\revision{Scheme of the proposed methodological framework:
				the RL algorithm learns a sensing design to maximize performance of the estimation algorithm.}}
		\label{fig:framework}
\end{figure}}

In contrast to previous work,
we jointly address sensor local processing,
computation and communication latency, 
and system dynamics \revision[2]{towards a dynamical smart} sensing design.

In~\cite{9137405},
the authors proposed a general model for a processing network,
including impact of computation-dependent delays on \revision{monitoring performance}, 
and provided a heuristic sensing design. 
However,
that design is static.
\ie sensors 
cannot adapt to the monitored system during operation,
which may hinder performance.
For example, time-varying systems 
generally prevent the optimal sensing configuration to be static.
Also, sensors could store incoming samples into an unlimited buffer.
We advance such issues through a novel design framework that builds on the insights in~\cite{9137405}.
Moreover,
this article considerably expands the preliminary version~\cite{cdc} as described next.

First, in~\autoref{sec:system-model} we propose a \revision{novel model} for a \emph{processing network}
\revision{tailored to data acquisition and transmission} by resource-constrained smart sensors.
These can adapt their local computation overtime and exploit the \tradeoff online to maximize global network performance,
by choosing to either transmit raw samples or refine data on-board.
In addition, 
motivated by~\cite{9137405},
we let sensors temporary stand-by (\textit{sleep}) to alleviate the computational burden for 
\revision{sensor fusion.
Roughly speaking,
such \textit{online sensor selection} can crucially improve global monitoring performance
if the processing resources available at the base station
cannot handle large amounts of sensory data in real time.}
\revision{Remarkably,
this result goes against the common wisdom that deploying more sensing resources always improve performance.}

In~\autoref{sec:problem-formulation},
we formulate an optimal design problem to manage \revision{sensing} resources in a network of smart sensors.
We do this by computing an estimation-theoretic performance metric
\revision{that embeds both dynamical parameters and
accuracy and delays associated with sensory data.}
To partially overcome intractability of the problem,
in~\autoref{sec:centralized-policy} we formulate a simplified version of it,
which is tackled via a Reinforcement Learning approach in~\autoref{sec:RL-formulation},
see~\autoref{fig:framework}.
Reinforcement Learning, and data-driven methods in general,
are now popular in network systems and edge computing
because of challenges raised by real-world scenarios~\cite{shi_artificial_2020,baggio_data-driven_2021,wang_when_2018}.

Finally, in~\autoref{sec:simulation} we validate our approach with \revision{numerical experiments}
motivated by sensing for autonomous driving and Internet-of-Drone tracking.
We address realistic communication
through an industrial-oriented simulator (\omnet)
that accurately models 
the lower layers of the protocol stack.
We show that accounting for \revision{latency due to resource constraints} can improve performance through a careful allocation of \revision{sensing and computation}.
\revision{In particular,
the online sensor selection becomes crucial when a large number of sensors is available.}

%% file: Sections/setup.tex

\section{Setup and Problem Formulation}\label{sec:setup}

In this section,
we first model a processing network composed of smart sensors (\autoref{sec:system-model}),
and then formulate the sensing design as an optimal estimation problem (\autoref{sec:problem-formulation}).

%% file: Sections/system-model.tex

\subsection{System Model}\label{sec:system-model}

\textbf{Dynamical System.}
The signal of interest is described by a time-varying discrete-time linear dynamical system,
\begin{equation}\label{eq:stateEquation}
	\x{k+1} = A_k\x{k} + \w{k},
\end{equation}
where $x_k\in\Real{n}$ collects the \revision{variables (\textit{state})} of the system,
$A_k\in\Real{n\times n}$ is the state matrix,
and white noise $\w{k}\sim\gauss(0,W_k)$ captures model uncertainty.
Such class of models is widely used in control applications,
by virtue of their simplicity but also powerful
expressiveness~\cite{medeiros_distributed_2008,yang_leveraging_2020,jeon_stealthy_2019,radisavljevic-gajic_vulnerabilities_2018,devos_sensor_2021}.
For example, a standard approach in control of systems modeled through nonlinear differential equations
is to approximate the original model as a parameter- or time-varying linear system,
for which efficient control techniques are known~\cite{devos_sensor_2021,gros_linear_2020,tsai_robust_2014}.

In view of 
transmission of sensor samples,
we assume discrete-time dynamics with time step $T$, where subscript
$k\in\mathbb{N}$ means the $ k $th time instant $ kT $.
Without loss of generality, 
we fix the first instant $ k_0 = 1 $.
The sampling time $T$ represents a suitable time scale for the global monitoring
and, possibly, decision-making task at hand. 
For example, typical values of $ T $ are one or two seconds for trajectory planning of ground robots,
while higher frequency is required for drones performing a fast pursuit or for self-driving applications. 

%% file: Sections/smart-sensors.tex

\textbf{Smart Sensors.}
The system modeled by~\eqref{eq:stateEquation} is measured by $N$ smart sensors (or simply {sensors}) 
gathered in the set $ \sensSet \doteq \{1,\dots,N\} $,
which output a noisy version of the state $ x_k $,
\begin{equation}\label{eq:sensorMeasurement}
	\yi{k}{i} = \x{k} + \vi{k}{i}, \qquad \vi{k}{i}\sim\gauss(0,\V{i,k}),
\end{equation}
where $\yi{k}{i}$ is the measurement \revision{produced}
by the $ i $th sensor at time $ k $,
for any $ i\in\sensSet $,
and $\vi{k}{i}$ is measurement noise.

Smart sensors \revision{are equipped with processing capabilities
	alongside standard sensing hardware,
	and can either transmit raw samples of the signal $ \x{k} $
	or locally process acquired samples to provide refined measurements.
	For example,
	a smart camera may
	send raw frames
	or run computer vision algorithms on the acquired images
	to get high-quality information,
	as in typical robot navigation applications
	where informative features are extracted from visual data.
	Symbol $ \yi{k}{i} $ indifferently refers to raw or processed measurements:
	as formalized in~\cref{ass:delays-variances},
	the difference between such two kinds of data is embedded into the measurement noise covariance $ \V{i,k} $.
	
	\begin{rem}[Sensor processing]
		We consider the case where sensor local processing is \textit{static},
		that is,
		sensors can refine the current sample
		(as in~\cite{lopez2020cdc_latencyScheduling,9137405}),
		but do not (re-)process past samples.
		This model is suitable to devices
		that provide data 
		without need (or possibility) of tracking the history of the measured signal,
		which is handled by the base station.
		This is different than,
		\eg works~\cite{hovareshti_scheduling_2007,gupta_erasure_2009},
		where sensor processing is adaptive and involves the history of collected samples.
		Although it might be possible to integrate such kind of processing into our framework,
		this is a compelling research direction that will be explored in the future.
	\end{rem}
}
Sensors face a \textit{\tradeoff} through limited hardware:
raw data are less accurate,
but local data processing introduces extra computational delays
\revision{that make refined updates more outdated with respect to the current state of the system}.

For example, 
consider a car that is approximately moving at constant speed,
with $ w_k $ capturing small unmodeled accelerations:
as the car moves,
knowledge of its real-time position through the nominal model (constant speed)
becomes more and more imprecise because of unknown accelerations hidden in $ w_k $,
which make the car drift away from its nominal trajectory.
In this case, a sensor may prefer to sample the system (\eg collect positions of the car) more often,
rather than spending time to obtain precise, 
but outdated, 
position measurements.

\revision{
	We formally model the \tradeoff
	with the next assumptions.
	We also introduce a third operating mode (\textit{sleep mode})
	that lets sensors stand-by.
	The usefulness of sleep mode 
	is associated with limited computational resources for aggregation of sensory data,
	and will be motivated in the paragraph ``Base Station'' below and in~\autoref{sec:problem-formulation}.}

\begin{ass}[Sensing modes]\label{ass:delays-variances}
	Each sensor $ i\in\sensSet $ can be in \textit{raw}, \textit{processing}, or \textit{sleep} mode.
	\begin{description}
		\item[Raw mode:] measurements are generated after {delay} $ \delayRaw[i] $ with noise covariance $\V{i,k}\equiv\varRaw[i]$. 
		\item[Processing mode:] measurements are generated after {\emph{processing delay}} $\delayProc[i] $ with noise covariance $\V{i,k}\equiv\varProc[i]$. 
		\item[Sleep mode:] the sensor is temporary set idle (\textit{asleep}): neither data sampling nor transmission occur under this mode.
	\end{description}
\end{ass}

\begin{ass}[Latency-accuracy trade-off]\label{ass:tradeoff}
	For each sensor $ i\in\sensSet $, it holds $ \delayProc[i] > \delayRaw[i] $ and $\varRaw[i]\succ\varProc[i]$.%
	\footnote{
		Even though L\"{o}wner order obeyed by covariance matrices is partial,
		we require it in our model so that the latency-accuracy trade-off is well defined.}
\end{ass}

\begin{figure}
	\centering
	\includegraphics[width=\linewidth]{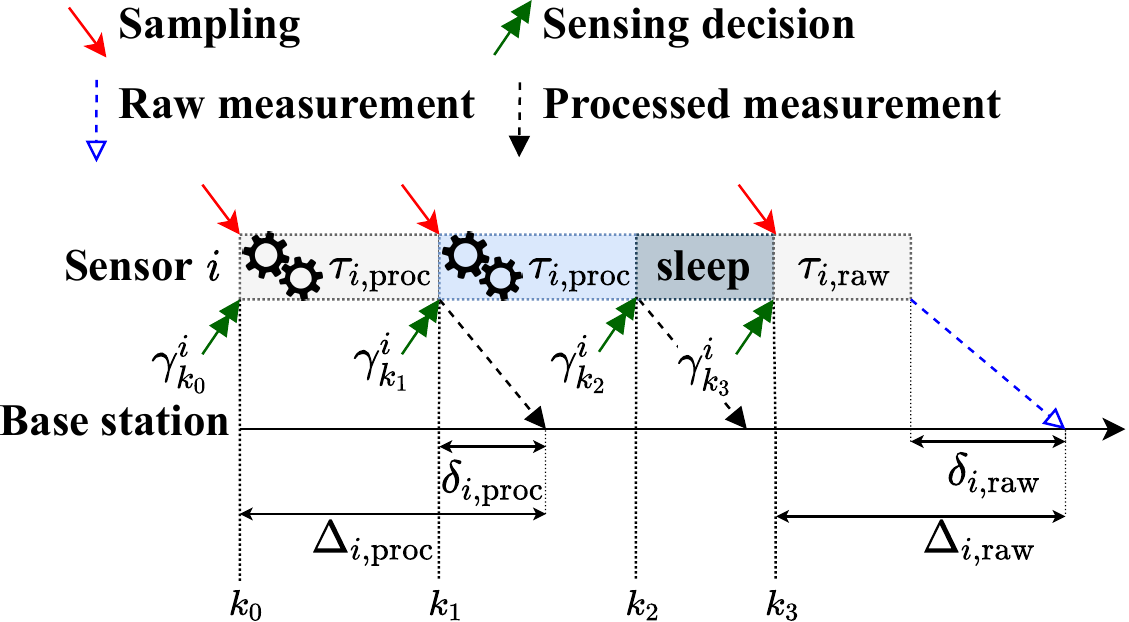}
	\caption{\textbf{Data collection and transmission.}
		Computation at the $ i $th sensor is ruled by sensing policy $ \policy[i] $.
		Here, sensing decisions
		$ \{\decision{k_j}{i}\}_{j=0}^3 = \{\proc,\proc,\sleep,\raw\}$ are shown
		and~\eqref{eq:sampling-sequence-subequations} reads
		$ \samplingSequence{i}[0] = s_i^0(k_0) = k_0 $, 
		$ \samplingSequence{i}[1] = s_i(k_0) = k_1 $, 
		and $ \samplingSequence{i}[2] = s_i(k_1) = k_3 $.
		Measurements are received after delays induced by local computation
		(rectangular blocks) and communication (dashed arrows).
		For example, under $ \decision{k_0}{i} = \proc $,
		the sample acquired at time $ k_0 $ is first processed 
		(with processing delay $ \delayProc[i] $),
		then transmitted at time $ k_1 = k_0 + \delayProc[i] $
		(with communication delay $ \delayCommProc[i] $),
		and finally received at the base station at time $ k_1 + \delayCommProc[i] = k_0 + \delayRecProc[i] $
		(with delay at reception $ \delayRecProc[i] $).
	}
	\label{fig:delayed-meas-reception}
\end{figure}

Similarly to~\cite{9137405},
\cref{ass:tradeoff} models high accuracy
through long computation (\autoref{fig:delayed-meas-reception}) and ``small'' covariance (intensity) of measurement noise,
\eg raw distance measurements with uncertainty of $ 1\si{\meter^2} $ 
while processed ones of $ 0.1\si{\meter^2} $.

Next, we define how local operations are ruled overtime.

\begin{definition}[Sensing policy]\label{def:processing-policy}
	A \emph{sensing policy} for the $ i $th sensor is a sequence of categorical decisions $ \policy[i] \doteq \{\decision{k}{i}\}_{k\ge k_0} $.
	If $ \decision{k}{i} = \raw $, measurement $ \yi{k}{i} $ is transmitted raw;
	if $ \decision{k}{i} = \proc $, $ \yi{k}{i} $ is processed;
	if $ \decision{k}{i} = \sleep $, no measurement is acquired at time $ k $.
\end{definition}

According to~\cref{def:processing-policy}, different sensing modes can be alternated online.
However, because of constrained resources,
a sensor cannot acquire measurements arbitrarily often.
The actual sampling frequency is determined
as formalized next. 

\begin{ass}[Sampling frequency]\label{ass:sensor-frequency}
	{Assume that the $ i $th sensor acquires a sample at time $ k $ 
		under either raw ($ \decision{k}{i} = \raw $) or processing ($ \decision{k}{i} = \proc $) mode.
		Let time $ k' $ be defined as
		\begin{subequations}\label{eq:sampling-instant-next}
			\begin{equation}\label{eq:sampling-instant-next-no-sleep}
				k' \doteq \begin{cases}
					k + \delayRaw[i]  										& \text{if }\decision{k}{i}=\raw\\
					k + \delayProc[i] 										& \text{if }\decision{k}{i}=\proc
				\end{cases}.
			\end{equation}
			Then, the next sample (under any mode) occurs at time $ s_i(k) $,
			\begin{equation}\label{eq:sampling-instant-next-general}
				s_i(k) \doteq \min_{h\in\mathbb{N}}\lb h \ge k':\decision{h}{i}\neq\sleep\rb.
			\end{equation}
		\end{subequations}
		Finally, the sequence of all sampling instants $ \samplingSequence{i} $ is given by
		\begin{subequations}\label{eq:sampling-sequence-subequations}
			\begin{equation}\label{eq:sampling-sequence}
				\samplingSequence{i} = \lb s_i^l(k_0) \rb_{l\ge0}, 
			\end{equation}
			where consecutive sampling times are defined by the recursion
			\begin{equation}\label{eq:sampling-instant-chain}
				\begin{aligned}
					s_i^{l+1}\left(k\right) &= s_i\lr s_i^i(k)\rr\\
					s_i^0\left(k_0\right) 	&\doteq \min_{h\in\mathbb{N}}\lb h \ge k_0:\decision{h}{i}\neq\sleep\rb,
				\end{aligned}
			\end{equation}
		\end{subequations}
		and $ \samplingSequence{i}[l] $ denotes the $ l $th element of the sequence, with $ l\in\mathbb{N} $.}
\end{ass}

In words,~\cref{ass:sensor-frequency} states that sensors
can acquire a new sample only after the previous measurement has been transmitted. 
This is a realistic assumption if agents have 
limited storage resources~\cite{mildenhall_limited_storage_2020}.
The effect of a sensing policy on sampling and local data processing is illustrated in~\autoref{fig:delayed-meas-reception}.

%% file: Sections/wireless-channel.tex

\textbf{\revision{Communication} channel.}
All sensors transmit data to a common base station through a shared \revision{communication channel,
which is wireless or wired according to the application requirements}.
The channel induces \emph{communication latency}
that may further delay transmitted updates
\revision{and depends on several factors such as transmission medium, 
	network traffic, 
	and interference.}
We let $ \delayCommRaw[i] $ and $ \delayCommProc[i] $ denote 
communication delays of raw and processed data transmitted by the $ i $th sensor,
respectively.
In general,
$ \delayCommRaw[i] $ and $ \delayCommProc[i] $ might differ
depending on possible data compression due to processing.
In case $ \delayCommRaw[i] = \delayCommProc[i] $,
we denote both delays by $ \delayComm[i] $.\linebreak
The total delay experienced by updates
from sampling to reception at the base station 
is given by $ \delayRecRaw[i] = \delayRaw[i] + \delayCommRaw[i]$ for raw
and $ \delayRecProc[i] = \delayProc[i] + \delayCommProc[i] $ for processed data.
Data sampling, processing,
and transmission are depicted in~\autoref{fig:delayed-meas-reception}.

%% file: Sections/base-station.tex

\begin{figure}
	\centering
	\includegraphics[width=0.8\linewidth]{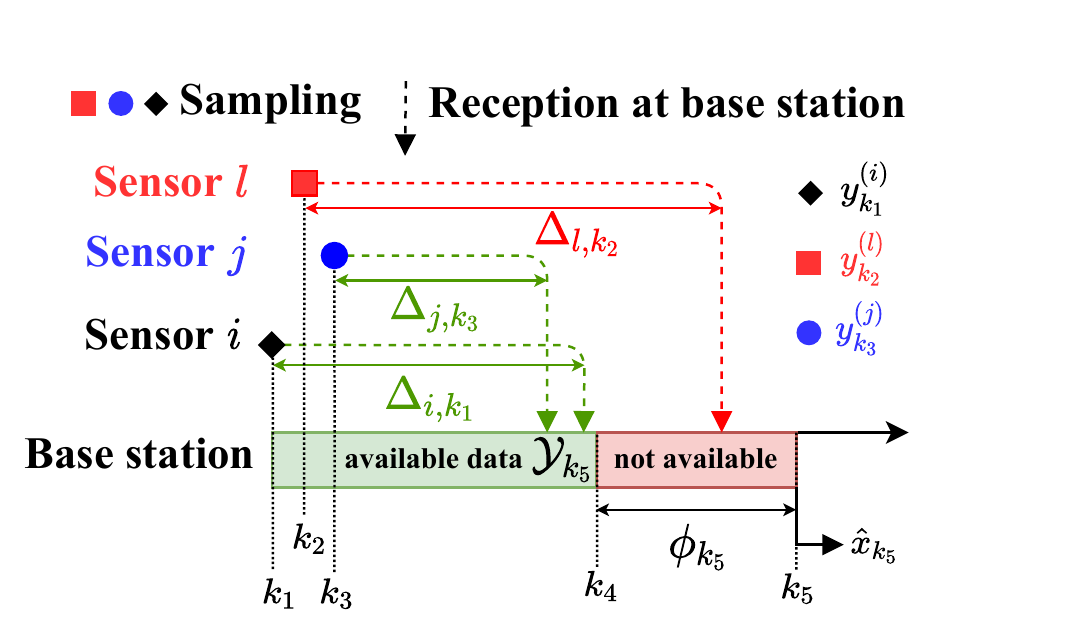}
	\caption{\textbf{Data processing at the base station.}
		Resource-constrained centralized processing introduces \emph{fusion delay} $ \delayFus{k_5} $
		to estimate $ \x{k_5}{} $. 
		Measurements $ \yi{k_1}{i} $ and $ \yi{k_3}{j} $ are received before computation starts at time $ k_4 = k_5 - \delayFus{k_5} $ 
		and are used to compute $ \xhat{k_5}{} $,
		\ie $ \yi{k_1}{i}, \yi{k_3}{j} \in\measurements{k_5} $,
		while $ \yi{k_2}{l} $ is received after time $ k_4 $ and cannot be used in estimation of $ \x{k_5} $, 
		\ie $ \yi{k_2}{l} \notin\measurements{k_5} $. 
	}
	\label{fig:delayed-meas-fusion}
\end{figure}

\textbf{Base Station.}
Data are transmitted 
to a base station in charge of estimating the state of the system $ \x{k} $ in real time.
Such estimation enables remote global monitoring and decision-making,
{\eg coordinated tracking or exploration}. 
Let $ \xhat{k}{} $ denote the real-time estimate of $ \x{k} $.
In view of the sequential nature of centralized data processing,
the real-time estimate of $ \x{k} $ is computed in $ \delayFus{k} $ time (\emph{fusion delay}),
which is proportional to the amount of data used in the update~\cite{9137405}. 
Consider~\autoref{fig:delayed-meas-fusion}:
from time $ k_1 $ through $ k_4 $, 
new data are received at the base station (green dashed arrows).
If the estimation routine starts at time $ k_4 $,
it takes $ \delayFus{k_5} $ to process all newly received sensory data
(possibly, also old ones if some data arrive out of sequence),
and hence the next updated state estimate, $ \xhat{k_5}{} $, will be available at time $ k_5 = k_4 + \delayFus{k_5} $.
Hence,
fusion delays induce open-loop predictions
that degrade quality of the computed estimates 
(similarly to what discussed about local sensor processing),
and motivate sleep mode to reduce the incoming stream of sensory data and improve overall performance~\cite{9137405}.  

\begin{ass}[Available sensory data]\label{ass:available-data}
	In view of Assumptions~\ref{ass:delays-variances},~\ref{ass:sensor-frequency},
	all sensory data available at the base station and used to compute $ \xhat{k}{} $ at time $ k $ are
	\begin{gather}
		\nonumber\measurements{k} \doteq {\bigcup_{i\in\sensSet}\bigcup_{l\in\mathbb{N}}}
		\lb \lr\yi{\samplingSequence{i}[l]}{i},\V{i,\samplingSequence{i}[l]}\rr : \samplingSequence{i}[l] + \delayRec{i,\samplingSequence{i}[l]} + \delayFus{k} \le k\rb \\
		\delayRec{i,\samplingSequence{i}[l]} \doteq \begin{cases}
			\delayRecRaw[i]  & \text{if }\decision{\samplingSequence{i}[l]}{i}=\raw\\
			\delayRecProc[i] & \text{if }\decision{\samplingSequence{i}[l]}{i}=\proc
		\end{cases}, \label{eq:sequence-measurements}
	\end{gather}
	where the $ l $th measurement from the $ i $th sensor $ \yi{\samplingSequence{i}\left[ l \right]}{i} $
	is sampled at time $ \samplingSequence{i}\left[ l \right] $ and
	received after overall delay $ \delayRec{i,\samplingSequence{i}\left[l\right]} $,
	and $ \delayFus{k} $ is the time needed to compute $ \xhat{k}{} $ at the base station.
\end{ass}
According to~\cref{ass:available-data},
a measurement $ \yi{h}{i} $ can be used to compute the estimate of $ \x{k} $ in real time 
if it is successfully delivered to the base station (with delay at reception $ \delayRec{i,h} $) before or at time $ k-\delayFus{k} $,
where $ \delayFus{k} $ is the amount of time needed to compute $ \xhat{k}{} $.
Data processing 
at the base station with limited resources and data availability
is depicted in~\autoref{fig:delayed-meas-fusion}.

\begin{figure}
	\centering
	\includegraphics[width=.8\linewidth]{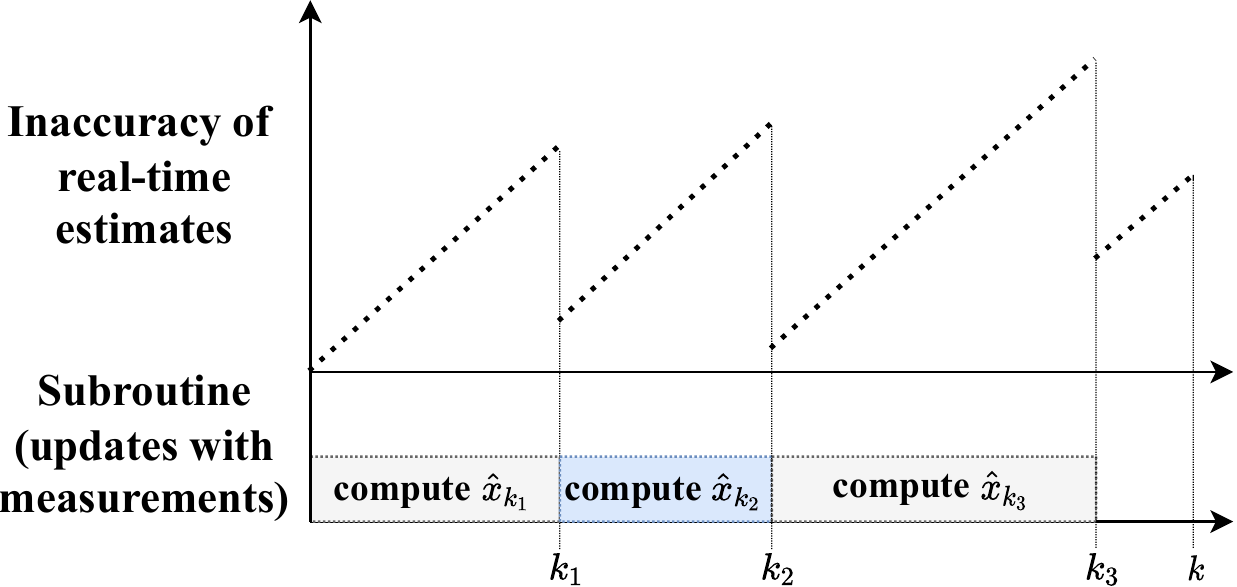}
	\caption{\textbf{Real-time estimation at the base station.}
		The state estimate is updated at each point in time (top).
		Because of limited resources at the base station,
		open-loop updates are performed whenever fresh sensory data are being processed (bottom),
		causing estimation to degrade overtime through additive noise $ \w{k} $ in nominal dynamics~\eqref{eq:stateEquation}.
		As soon as the data processing subroutine produces an updated estimate with new measurements,
		\eg $ \xhat{k_1}{} $ at time $ k_1 $,
		the estimation inaccuracy is reduced. 
		Note that the top plot is qualitative:
		the estimate quality does not degrade linearly, in general.
	}
	\label{fig:estimation-routines}
\end{figure}

\begin{rem}[Real-time estimation]
	Based on the above discussion,
	new data cannot be used by an estimation procedure between times $ k $ and $ k + \delayFus{k} $.
	In a real system,
	a real-time state estimate must always be available for effective monitoring.
	We assume that two parallel jobs are executed.
	A support subroutine processes received measurements
	and computes a state estimate at time $ k $ in $ \delayFus{k} $ time (\cf~\autoref{fig:delayed-meas-fusion}). 
	The real-time estimation routine computes one-step-ahead open-loop updates at each point in time
	according to the nominal dynamics~\eqref{eq:stateEquation} 
	(progressively degrading estimation quality),
	and resets when the support subroutine outputs an updated estimate with new measurements 
	(with higher estimate quality).\footnote{
	One-step-ahead open-loop steps are assumed computationally cheap.}
	A schematic representation is shown in~\autoref{fig:estimation-routines}.
	Importantly, degradation of estimation in the top plot
	is not due to lack of new measurements (like in Age of Information literature),
	but is caused by constrained resources that induce a computational bottleneck in the support subroutine
	(bottom plot in~\autoref{fig:estimation-routines}).
\end{rem}


%% file: Sections/problem-formulation.tex

\subsection{Problem Statement}\label{sec:problem-formulation}

The trade-offs introduced in the previous section
call for a challenging sensing design at the network level. 
In particular, 
all possible choices of local sensor processing 
(we address a specific choice for all sensors as a \textit{sensing configuration})
affect global performance in a complex manner,
whereby it is unclear
which sensors should
transmit raw measurements,
with poor accuracy and possibly long communication delays,
and which ones should refine their samples locally
to produce high-quality measurements.
In fact, the authors in~\cite{9137405} show that 
the optimal configuration when considering steady-state performance is nontrivial.
Also, the 
optimal sensing configuration is time varying, in general. 
Thus, 
sensing policies $ \policy[i], i\in\sensSet $, have to be suitably designed
to maximize the overall network performance.

The state $ \x{k} $ 
is estimated via Kalman predictor,
which is the optimal observer for linear systems with Gaussian disturbances.
It can be shown,
\eg via state augmentation,
that the Kalman predictor is optimal even with delayed measurements,
whereby it suffices to ignore updates 
associated with missing data (see Appendix~\ref{app:kalman-filter} in Supplementary Material).
Out-of-sequence arrivals
can be handled by recomputing all predictor steps since the latest arrived measurement has been acquired,
or by more sophisticated techniques~\cite{kalmanDelays,kalmanDelays1}.

Let $\xtilde{k}{}\doteq\x{k}-\xhat{k}{}$ the estimation error of Kalman predictor at time $ k $, 
and let $ \Pmat{k}{}\doteq\var{\xtilde{k}{}} $ its covariance matrix.
We formulate the sensing design as an optimal estimation problem.

\begin{prob}[Sensing Design for Processing Network]\label{prob:optimal-estimation}
	Given system~\eqref{eq:stateEquation}--\eqref{eq:sensorMeasurement} and Assumptions~\ref{ass:delays-variances}--\ref{ass:available-data},
	find the optimal sensing policies $ \policy[i] $, $ i\in\sensSet $,
	that minimize the time-averaged estimation error variance with horizon $ K $,
	\begin{argmini!}
		{\substack{\policy[i]\in\Pi_i,i\in\sensSet}}
		{\frac{1}{K}\sum_{k=k_0}^K \tr{\Pmat{k}{\policy}}\protect\label{eq:prob-optimal-estimation-objective}}
		{\label{eq:prob-optimal-estimation}}
		{}
		\addConstraint{\Pmat{k}{\policy}}{= \kalman{\measurements[\policy]{k}}\protect\label{eq:prob-optimal-estimation-constraint-kalman}}
		\addConstraint{\Pmat{k_0}{\policy}}{= P_0\protect\label{eq:prob-optimal-estimation-constraint-init-cond}},
	\end{argmini!}
	where the Kalman predictor $ \kalman{\cdot} $ computes at time $ k $ the state estimate $ \xhat{k}{\policy} $ 
	and the error covariance matrix $ \Pmat{k}{\policy} $
	using data $ \measurements[\policy]{k} $ available at the base station according to $ \policy \doteq \{\policy[i]\}_{i\in\sensSet} $,
	and $ \Pi_i $ gathers all causal sensing policies of the $ i $th sensor.
\end{prob}

\revision{\begin{rem}[Impact of processing on estimation]
	Processed measurements are more accurate than raw ones:
	hence,
	if delays were neglected,
	the optimal (trivial) design would be to always process,
	because this yields the smallest variance of measurement noise (\cref{ass:delays-variances})
	and minimizes the estimation error variance of the Kalman predictor
	when updates with measurements are performed.
	However,
	computational delays associated with data processing introduce extra
	open-loop steps that increase the error variance,
	making the optimal design nontrivial.
	In other words,
	uncertainty about the true dynamics
	(captured in~\eqref{eq:stateEquation} by noise $ w_k $)
	makes refined measurements be less informative about the \textit{current} state of the system,
	so that high accuracy alone might not pay off in real-time monitoring.
\end{rem}}

\begin{rem}[\revision{Novelty of} sensor selection]\label{rem:difference-w-sensor-selection}
	Sleep mode actually implements an \textit{online sensor selection},
	\revision{whereby sleeping sensors do not supply data to the base station.}
	We identify two key elements that make our framework fundamentally different
	from standard sensor selection in the literature.
	First, 
	\revision{while we exploit sleep mode
	towards \emph{optimal performance}},
	sensors are typically selected \revision{to trade performance for available resources
	under the conventional belief that more sensors
	yield better performance.}
	In contrast, 
	selection emerges \emph{naturally} in our framework
	\revision{\textit{to maximize performance}}
	in view of the computational bottleneck at the base station
	that may increase the estimation cost in~\eqref{eq:prob-optimal-estimation}.
	Moreover,
	rather than a \emph{static} selection,
	we allow for \emph{dynamical switching} to and from sleep mode,
	which both enables performance improvement through richer design options
	\revision{and is more challenging to optimize}.
\end{rem}

%% file: Sections/centralized-implementation.tex

\section{\titlecap{sensing policy: a centralized implementation}}\label{sec:centralized-policy}

\cref{prob:optimal-estimation} is combinatorial \revision{in the number of sensors}
and 
raises a computational challenge in finding efficient sensing policies,
because the search space may easily explode. 
For example,
$ 10 $ sensors yield $ 2^{10} = 1024 $ possible sensing configurations at each sampling instant.
Also,
\cref{prob:optimal-estimation} requires to design 
\revision{a potentially asynchronous schedule for each sensor,
	which is an additional combinatorial problem in the time horizon $ K $.
	To further complicate things,
	a sensing policy $ \policy[i] $ not only affects delay and accuracy of measurements supplied by the $ i $th sensor,
	but also determines the very sequence of sampling instants $ \samplingSequence{i} $
	(\cf~\eqref{eq:sampling-instant-next}--\ref{eq:sampling-sequence-subequations}),
	augmenting the search space to all possible time sequences over $ K $ steps.
	In particular,
	sleep mode represents a computational challenge,
	because it requires evaluation of all instants subsequent to its activation 
	to decide the best time for triggering a new update.
}

To partially ease the intractability of the problem, 
and motivated by practical applications,
we restrict the domain of candidate sensing policies
to reduce problem complexity
while maintaining a meaningful setup. 
First,
we look at the simple but relevant scenario with a homogeneous network
and motivate the design of a centralized policy in~\autoref{sec:hom-sens}.
We then go back to the general scenario with a heterogeneous network
and formulate a simplified version of~\cref{prob:optimal-estimation}
in~\autoref{sec:het-sens}.

%% file: Sections/homogeneous-sensors.tex

\subsection{Homogeneous Network}\label{sec:hom-sens}

\textbf{Sensor Model.}
In this scenario, all smart sensors have equal measurement noise distributions,
\begin{equation}\label{eq:hom-sensors}
    \yi{k}{i} = \x{k} + \vi{k}{i}, \qquad \vi{k}{i}\sim\gauss(0,\V{k}),
\end{equation}
with 
$ \V{k} = \varRaw $ or $ \V{k} = \varProc $ for raw and processed data, respectively.
Also, all sensors feature identical computational and transmission resources, 
given by delays $ \delayRaw $, $ \delayCommRaw $ for raw measurements
and $ \delayProc $, $ \delayCommProc $ for processed measurements, respectively
($ \delayComm $ in case of no compression). 
This \textit{homogeneous network}
models the special but relevant case where sensors are interchangeable.
This happens for example with sensor networks measuring
temperature in plants or chemical concentrations in reactors. 
Also, this model captures smart sensors collecting high-level environmental information,
such as UAVs tracking the position of a body moving in space. 


\textbf{Centralized Policy.}
In this case, 
it is sufficient to decide \textit{how many}, rather than \emph{which},
sensors follow a certain mode. 
Accordingly,
we focus on the design 
of a centralized policy that commands all sensors with no distinctions among them. 

\begin{definition}[Homogeneous sensing policy]\label{def:hom-sens-policy}
	A \emph{\homProcPol} is a sequence of categorical decisions $ \policyHom = \{\decisionHom{\ell}\}_{\ell=1}^L $.
	Each decision $ \decisionHom{\ell} = (N-\ns,\np) $ is taken at time $ \kell[]{} $
	such that
	$ \ns $ sensors are in sleep mode and
	$ \np $ out of the other $ N-\ns $ sensors are in processing mode
	between times $ \kell[]{} $ and $ \kell[+1]{} $,
	with $ 0\le\ns\le N $ and $ 0\le\np\le N-\ns $.
	Without loss of generality, we set $ k^{(1)} = k_0 $, $ \kL \le K $.
\end{definition}

\begin{figure}
	\centering
	\includegraphics[width=\linewidth]{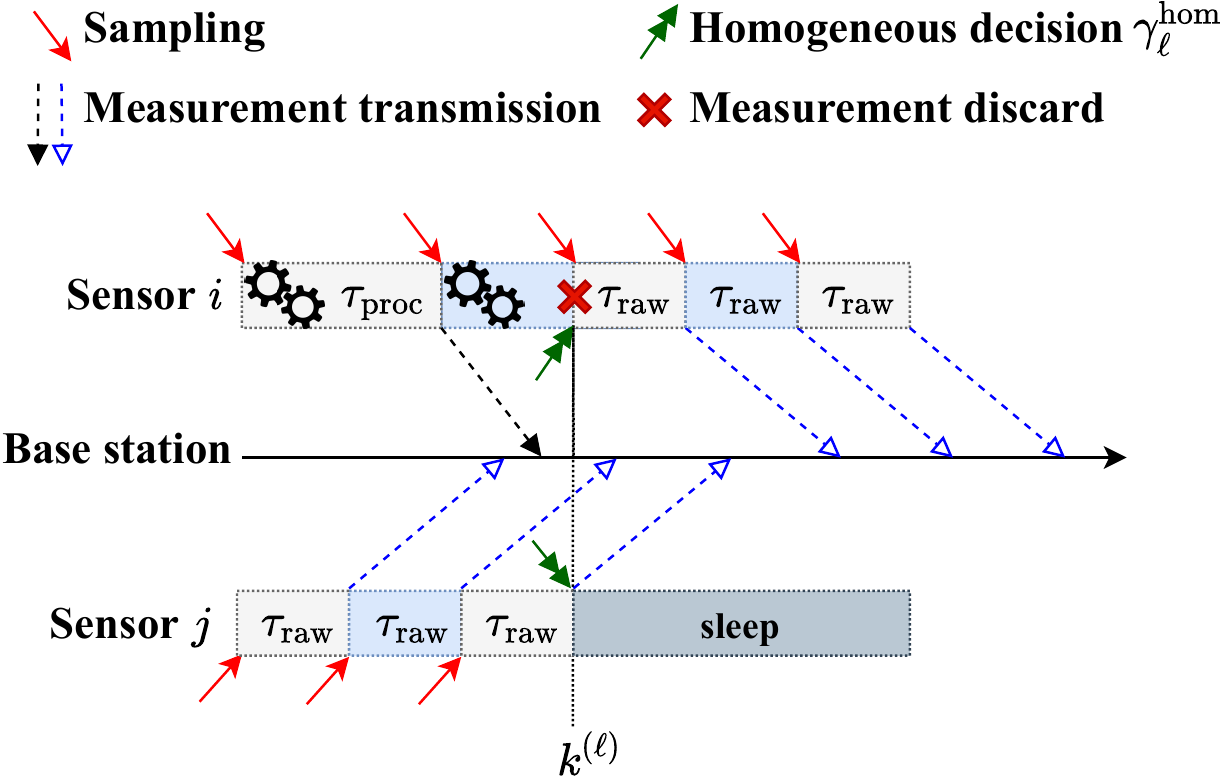}
	\caption{\textbf{Homogeneous sensing policy.} 
		Sampling and data processing at identical sensors are ruled by policy $ \policyHom $.
		Decision $ \decisionHom{\ell} $ is communicated at time $ \kell{} $
		and realized at individual sensors as $ \decision{\ell}{i} = \raw $ and $ \decision{\ell}{j} = \sleep $. 
		Concurrently, the $ i $th sensor disregards its current processed measurement (red cross) and
		switches to raw mode, acquiring a new sample at time  $ \kell{} $.
	}
	\label{fig:delayed-meas-centr}
\end{figure}

In words, the base station decides a configurations for all sensors at predefined time instants, 
which is both practical for applications
and convenient to reduce complexity of the problem. 
However, decisions may be taken at any times,
as long as these are consistent with sensor computational delays
(\eg to guarantee that one sample is collected for each decision).

With a slight abuse of notation, 
to denote the mode of a specific sensor that is
following the homogeneous decision $ \decisionHom{\ell} $, 
we write \revision{$ \decision{\ell}{i} = \mathrm{m} $
meaning that the $ i $th sensor is set in mode $ \mathrm{m} $ by the $ \ell $th homogeneous decision,
where $ \mathrm{m}\in\{\raw,\proc,\sleep\} $ can be 
raw ($ \raw $),
processing ($ \proc $),
or sleep mode ($ \sleep $),
respectively.}
We stress that in this context
$  \decision{\ell}{i} $ does not represent a decision of a \emph{single-sensor} sensing policy $ \policy[i] $ (as in~\cref{def:processing-policy}),
but all decisions are centralized
and $ \decision{\ell}{i} $ denotes the mode that \emph{the base station commands} to the $ i $th sensor through decision $ \decisionHom{\ell} $. 

By design,
centralized decisions are communicated regardless of current sensing status. 
In light of common practice in real-time control~\cite{5510096,9408050,sym12010172,s22041638,8950356},
we assume what follows.

\begin{ass}[Sampling frequency with \homProcPol]\label{ass:decision-time-centralized-policy}
	Decision $ \decisionHom{\ell} $ switches mode of the minimum amount of sensors possible.
	If the $ i $th sensor switches mode,
	the measurement currently being acquired or processed (if any) is immediately discarded.
	If the new commanded mode is either raw or processing,
	a new sample is acquired according to such new mode right after the decision $ \decisionHom{\ell} $ is communicated.
	
	Formally, given measurement $ \yi{k}{i} $ sampled at time $ k < \kell{} $
	obeying decision $ \decisionHom{\ell-1} $,
	the sampling dynamics~\eqref{eq:sampling-instant-next-general} becomes
	\begin{subequations}\label{eq:sampling-instant-next-centralized-policy-all}
		\begin{equation}\label{eq:sampling-instant-next-centralized-policy}
			s_i^\text{hom}(k)\doteq\begin{cases}
				k'			& \text{if }k^{(\ell)} \ge k'\\
				k^{(\bar{\ell})} & \text{otherwise},
			\end{cases}
		\end{equation}
		\begin{equation}\label{eq:l'}
			\bar{\ell}\doteq\min_{\ell'\in\{1,\dots,L\}}\left\lbrace \ell' : \ell'\ge \ell\wedge\decision{\ell'}{i}\neq\sleep\right\rbrace.
		\end{equation}
	\end{subequations}
	Further, $ \yi{k}{i} $ 
	is discarded (not transmitted) if 
	$ s_i^\text{hom}(k) \neq k' $.
\end{ass}

The new sampling mechanism is depicted in~\autoref{fig:delayed-meas-centr}.
According to~\cref{ass:decision-time-centralized-policy},
a measurement is not transmitted to the base station 
if it is not ready when a concurrent decision is communicated. 
In~\autoref{fig:delayed-meas-centr} the $ i $th sensor 
discards a measurement whose processing is not completed at time $ \kell{} $, 
when a new decision switches its mode.
Formally,
a sensor disregards raw (resp. processed) measurements sampled at time $ \bar{k} < \kell{} $
such that $ \bar{k} + \delayRaw > \kell{} $ ($ \bar{k} + \delayProc > \kell{} $),
\ie their acquisition ends after a \textit{different} mode is imposed by decision $ \decisionHom{\ell} $ (\cf~\eqref{eq:sampling-instant-next-centralized-policy}).
We denote by $ \measurements[\policyHom]{k} $
all available data at the base station at time $ k $ according to~\eqref{eq:sampling-instant-next-centralized-policy-all} and
such discard mechanism imposed by policy $ \policyHom $,
that excludes some data included in $ \measurements[]{k} $ (\cf~\eqref{eq:sequence-measurements}).

%% file: Sections/heterogeneous-sensors.tex

\subsection{Heterogeneous Network}\label{sec:het-sens}

We now return to the original model~\eqref{eq:sensorMeasurement} with heterogeneous sensors.
Without loss of generality, 
we assume that the sensor set $ \sensSet $ is partitioned into $ M $ subsets
$ \sensSet_1,\dots,\sensSet_M $,
where subset $ \sensSet_m $, $ m\in\{1,\dots,M\} $,
is composed of homogeneous sensors of the $ m $th class.
From what discussed in the previous section,
it is sufficient to specify how many sensors follow a certain mode within each subset $ \sensSet_m $.
Hence, 
we narrow down the  
domain of all possible policies 
according to the next definition.

\begin{definition}[Network sensing policy]\label{def:het-net-sens-pol}
	A \emph{\netProcPol} is a collection
	$ \policyNet \doteq \{\policyHom[m]\}_{m=1}^M $,
	where each \homProcPol $ \policyHom[m] $ is associated with homogeneous sensor subset $ \sensSet_m $,
	and all homogeneous decisions $ \{\decisionHom[m]{\ell}\}_{m=1}^M $ are 
	communicated together at time $ \kell[]{} $.
\end{definition}

In~\cref{def:het-net-sens-pol},
decision times are fixed like in the homogeneous case,
so that 
decisions are communicated to all sensors at once.
At time $ \kell{} $, homogeneous decision $ \decisionHom[m]{\ell} $ involves sensors in $ \sensSet_m $,
and the overall sensing configuration is given by the ensemble of such decisions.
All data available at the base station at time $ k $ are collected in
$ \measurements[\policyNet]{k} \doteq \{\measurements[\pi_{\text{hom},m}]{k}\}_{m=1}^M $.

Finally, we get the following simplified problem formulation.

\begin{prob}[Centralized Sensing Design for Processing Network]\label{prob:optimal-estimation-simplified}
	Given system~\eqref{eq:stateEquation}--\eqref{eq:sensorMeasurement} with Assumptions~\ref{ass:delays-variances}--\ref{ass:decision-time-centralized-policy},
	find the optimal \netProcPol $ \policyNet $ that minimizes the time-averaged estimation error variance with horizon $ K $,
	\begin{argmini!}<b>
		{\substack{\policyNet\in\Pi_\text{net}}}
		{\frac{1}{K}\sum_{k=k_0}^K \tr{\Pmat{k}{\policyNet}}\protect\label{eq:prob-optimal-estimation-simplified-objective}}
		{\label{eq:prob-optimal-estimation-simplified}}
		{}
		\addConstraint{\Pmat{k}{\policyNet}}{= \kalman{\measurements[\policyNet]{k}}\protect\label{eq:prob-optimal-estimation-simplified-kalman}}
		\addConstraint{\Pmat{k_0}{\policyNet}}{= P_0\protect\label{eq:prob-optimal-estimation-simplified-init-cond}},
	\end{argmini!}
	where the Kalman predictor $ \kalman{\cdot} $ computes at time $ k $ the state estimate $ \xhat{k}{\policyNet} $ and 
	the error covariance matrix $ \Pmat{k}{\policyNet} $,
	using data available at the base station according to $ \policyNet $,
	and $ \Pi_\text{net} $ is the space of causal network sensing policies.
\end{prob}



%% file: Sections/RL-formulation.tex

\section{Reinforcement Learning \revision[2]{Algorithm}}\label{sec:RL-formulation}

\revision{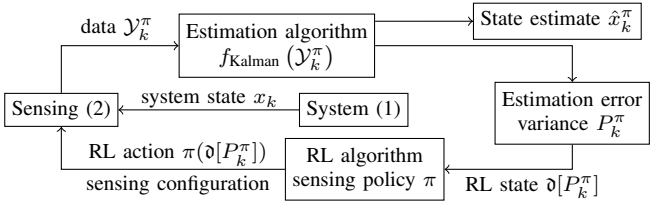
\begin{figure}[t]
	\centering
	\begin{tikzpicture}
		\begin{scope}[shift={(0,0)},scale=1.1]
			\footnotesize
			\node [draw] (sys) at (1,0) {Sensing~\eqref{eq:sensorMeasurement}};
			\node [draw, right=2.4cm of sys] (dyn) {System~\eqref{eq:stateEquation}};
			\path [->] (dyn) edge (sys);
			\node [above right=-.3cm and .2cm of sys] (sys-state) {system state $\x{k}$};
			\node [above right=.6cm and -.6cm of sys] (meas) {data $\measurements[\pi]{k}$};
			\node [draw, above right=.2cm and .8cm of sys] (kf) {\makecell{Estimation algorithm \\ $\kalman{\measurements[\pi]{k}}$}};
			\draw [->, to path={|- (\tikztotarget)}] (sys) edge (kf);
			\node [draw, above right= -.35cm and 1.3cm of kf] (est) {State estimate $\xhat{k}{\pi}$};
			\draw [<-] (est) edge (est -| kf.east);
			\node [draw, below right=-.6cm and 5cm of sys] (cov) {\makecell{Estimation error \\ variance $\Pmat{k}{\pi}$}};
			\draw [->, to path={-| (\tikztotarget)}] (kf) edge (cov);
			\node [draw, below right=.8cm and -1.2cm of kf] (rl) {\makecell{RL algorithm \\ sensing policy $\pi$}};
			\draw [->, to path={|- (\tikztotarget)}] (cov) edge (rl);
			\node [below left=.3cm and -1.5cm of cov] (state) {RL state $\mathfrak{d}[\Pmat{k}{\pi}]$};
			\draw [->, to path={-| (\tikztotarget)}] (rl) edge (sys);
			\node [left=0.1cm of rl] (sens) {\makecell{RL action $\pi(\mathfrak{d}[\Pmat{k}{\pi}])$\vspace{1mm}\\ sensing configuration}};
		\end{scope}
	\end{tikzpicture}
	\caption{\revision{\textbf{Learning framework.}
		The RL algorithm receives accuracy of estimates (state)
		and outputs sensing configurations (action) that affect sensory data.}}
	\label{fig:rl-scheme}
\end{figure}}

By assuming complete knowledge of delays 
and measurement noise covariances affecting sensors in the different modes,
both~\cref{prob:optimal-estimation} and~\ref{prob:optimal-estimation-simplified} can be analytically solved.
However, the computation of the exact minimizer requires to keep track of all starts and stops of data transmissions for each sensor,
resulting in a cumbersome procedure which admits no closed-form expression,
and requires to solve a combinatorial problem which does not scale with the number of sensors. 
Moreover,
the assumptions considered in the formulation of the problem may be too conservative in real-life scenarios,
and the latter method cannot be relaxed. 
Indeed, as long as either delays or covariances are not explicitly known or have some variability,
\ie they can be modeled by proper random variables,
the minimization becomes intractable.
This is true even if the expectations of these random variables are known,
since the dynamics in~\cref{prob:optimal-estimation-simplified} lead to a nonlinear behavior for the quantity to be minimized.

\revision[2]{For the reasons above, 
we tackle the problem of choosing the optimal sensing policy to minimize estimation uncertainty through a Reinforcement Learning (RL) algorithm,
which executes a sequential decision-making suitable for a dynamical sensing design and can flexibly address the general problem formulation.
Specifically,
the RL algorithm is run at the base station 
and implements a \netProcPol $\policyNet$ by iteratively choosing a sensing configuration at each time $\kell{}$.
A scheme of the overall framework is given in~\autoref{fig:rl-scheme}.}

%% file: Sections/RL-tradeoff.tex

\subsection{Optimizing Latency-Accuracy Trade-off}\label{sec:RL-tradeoff}

\revision{The Reinforcement Learning problem of maximizing a reward function through the correct sequence of actions is addressed in this work by the Q-learning algorithm. 
	The latter is a \textit{model-free} and \textit{off-policy} algorithm which updates the current estimate of the action-value-function targeting an \textit{optimistic} variant of the temporal-difference error. 
	In a finite Markov Decision Process, 
	this approach converges to the \emph{optimal} action-value function under standard Monro-Robbins conditions~\cite{NIPS1993_5807a685}, 
	and is efficient with respect to standard competitors~\cite{NEURIPS2018_d3b1fb02, pmlr-v139-li21b}.}

With regard to~\cref{prob:optimal-estimation}, 
{policy $\policy[i]$ is composed of categorical variables corresponding to sensing modes,
and characterizes the potential for intervention in the operations of the $ i $th sensor. 
The constraints due to the centralized implementation in~\cref{prob:optimal-estimation-simplified} 
allow us to consider 
a single policy $ \policyNet : \mathcal{S} \rightarrow \mathcal{A} $
describing how many sensors 
are required to process or sleep for each subset $ \sensSet_m $. 
In particular,
action $ a\in\mathcal{A} $ 
\revision{is described by $ M $ pairs of integers
specifying,
for each group $ \sensSet_m $,
(i) how many sensors
transmit and (ii) how many out of the latter ones
are in processing mode (\cf~\cref{def:hom-sens-policy})}.
For example,
if $ \sensSet = \sensSet_1 \cup \sensSet_2 $ with $ |\sensSet_1| = 4 $ and $ |\sensSet_2| = 5 $,
action $ a = \{(2,1),(3,0)\} $
means that\revision{,
within $ \sensSet_1 $,}
two sensors \revision{are commanded to} transmit,
\revision{one of these being} in processing mode,
and the other two sleep,
and similarly for $ \sensSet_2 $.

Since we aim to minimize the time-averaged error variance~\eqref{eq:prob-optimal-estimation-simplified-objective}, 
a straightforward metric to be chosen as reward function is the negative trace of matrix $\Pmat{k}{\policyNet}$, 
which evolves according to the Kalman predictor with delayed updates.
In the considered framework the base station is allowed to change sensing configuration (corresponding to a new action) at each time $\kell[]{}$,
therefore a natural way of defining the reward is to take the average of the negative trace of the covariance
during the interval between times $\kell[]{}$ and $\kell[+1]{}$,
so that the base station can appreciate the performance of a particular sensing configuration in that interval. 

This leads to the following instantiation of 
\revision{the RL problem,}
\begin{maxi!}<b>
	{\substack{\policyNet\in\Pi_\text{net}}}
	{-\mathbb{E}\ls\sum_{\ell=1}^{L}
		\frac{\gamma^\ell}{\kell[+1]{} - \kell{}}\sum_{k=\kell{}}^{\kell[+1]{}}\tr{\Pmat{k}{\policyNet}}\rs
		\protect\label{eq:RL-cost-specialized-objective}}
	{\label{eq:RL-cost-specialized}}
	{}
	\addConstraint{\Pmat{k}{\policyNet}}{= \kalman{\measurements[\policy]{k}}\protect\label{eq:RL-cost-specialized-kalman}}
	\addConstraint{\Pmat{k_0}{\policyNet}}{= P_0\protect\label{eq:RL-cost-specialized-init-cond}},
\end{maxi!}
with $ k^{(L+1)} \doteq K $.
The quantity of interest is the trace of the error covariance and thus a straightforward approach would take $\mathcal{S} = \mathbb{R}^+$.
To keep the {Q-learning} in a tabular \revision{(finite)} setting,
we discretize the state space 
through a function $\mathfrak{d}: \mathbb{R}^+\rightarrow\mathbb{N}^+$.
In particular,
the image of $\mathfrak{d}\ls\cdot\rs$ is given by $ M $ bins,
which were manually tuned in our numerical experiments to yield 
a fair representation of the values of $ \Pmat{k}{\policyNet} $ observed along the episodes.
Then, based on the bin associated with $\mathrm{Tr}(\Pmat{\kell{}}{})$,
the agent outputs a sensing configuration $ a\in\mathcal{A} $ through $\policyNet\lr\cdot\rr$ at each time $\kell{}$,
given by $a_\ell = \policyNet(\mathfrak{d}[\mathrm{Tr}(\Pmat{\kell{}}{})])$.

Notably, choosing $ \discFact = 1 $ and time intervals $ [\kell{},\kell[+1]{}] $ of equal length matches~\eqref{eq:RL-cost-specialized} and
the objective cost~\eqref{eq:prob-optimal-estimation-simplified-objective} exactly.}

\revision{
\begin{rem}[System dynamics and computational complexity]
    The Reinforcement Learning procedure is concerned only with the selection of the sensing scheme
    and not with computation of the estimate $ \xhat{k}{} $
    (contrary to data-driven estimation). 
    In particular, 
    the sensor configuration is optimized with respect to the evolution~\eqref{eq:RL-cost-specialized-kalman} of the covariance matrix $ \Pmat{k}{} $
    induced by the Kalman predictor 
    and not with respect to the actual system dynamics~\eqref{eq:stateEquation},
    which are measured by the sensors. 
    The Reinforcement Learning step is then independent from the dimension $ n $ of the original system~\eqref{eq:stateEquation}, 
    because it deals only with the error variance of state estimates,
    while the 
    estimation is performed by the Kalman predictor in a model-based fashion.
\end{rem}}

%% file: Sections/RL-discussion.tex

\revision{
	\subsection{Discussion: Challenges \revision[2]{of Reinforcement Learning}}\label{sec:RL-discussion}
	While the proposed RL-based solution can tackle~\cref{prob:optimal-estimation-simplified}
	in more flexible and efficient way than brute-force or greedy search,
	it also comes with nontrivial limitations
	\revision[2]{due to computational and performance challenges of RL algorithms},
	which we discuss next. 
	\revision[2]{Tackling such challenges requires dedicated efforts that will be addressed in follow-up work.}
	Note however that the framework considered in this article is representative of a broad range of control applications,
	and the following issues do not constitute a threat in the current setting.
	
	First,
	while we consider a single processing mode in~\cref{ass:delays-variances},
	a smart sensor may in general choose among several options to refine raw samples.
	For example,
	a robot equipped with cameras may run multiple geometric inference algorithms for perception,
	each trading runtime for accuracy~\cite{lopez2020cdc_latencyScheduling,chinchali2021network}.
	In general,
	the sensing policy of each sensor might feature several design options (modes),
	which in turn imply a larger action space for the Q-learning
	and might raise a nontrivial computational challenge
	because the total number of actions does not scale with the number of sensors.
	
	Second,
	even though we focus on a centralized learning technique,
	this inevitably leads to poor computational scalability \revision[2]{especially with heterogeneous sensors},
	because the action space is the combination of actions of individual sensors.
	While it is worth pointing out that many control and robotic applications involve either \textit{identical}
	or \textit{a few different} sensors,
	for which a centralized learning approach is feasible,
	investigating computationally efficient strategies to improve the scalability of the training in general is a relevant research question.
	One way to tackle this challenge might be Multi-Agent Reinforcement Learning~\cite{4445757},
	where each agent (here, smart sensor) receives the reward from the environment and autonomously trains its own policy,
	possibly exchanging information with other agents.
	In this scenario,
	each agent chooses only its own actions,
	so that the total number of actions actually scales with the number of agents and permits a computationally scalable training.
	Another argument in favor of this scenario is the possibility of training \textit{asynchronous} sensing policies tailored to the general problem formulation~\eqref{eq:prob-optimal-estimation},
	which is hardly solvable via centralized learning and might turn especially useful to effectively trigger the sleep mode.
	However,
	the price to pay is the reduced or absent coordination among the agents,
	which can slow down the overall training or even prevent convergence.
	
	Last, although the Q-learning algorithm is one of the most widespread algorithms because of its effectiveness and ease of implementation, the proposed procedure could be improved by refining some aspects of the current set up. One of the most challenging aspects is the handling of the continuous state-space, which has been solved through a simple discretisation. The latter can be seen as the simplest instance of function approximation, so there could be better ways of addressing the specific state-space resulting from the present formulation. It is nonetheless remarkable that satisfactory results can already be achieved with this simple version, proving the flexibility of the Q-learning algorithm. Note indeed that the particular framework of a continuous state-space and a discrete action-space significantly reduces the range of algorithms that can be applied to solve the problem addressed in this work. 
	\revision[2]{The Q-learning algorithm proved flexible enough to handle the difficulties of the non-Markovian environments in both settings considered in~\autoref{sec:simulation}.
		To assess its effectiveness, the convergence is numerically studied in Appendix~\ref{app:convergence} in the Supplementary Material,
		together with a discussion on how the sample complexity scales with the number of sensors in the homogeneous setting. 
		While more powerful RL algorithms 
		could give better solutions, an extensive investigation is out of the scope of this work, whose goal is to propose a general methodology for the addressed sensing design problem.
	}
}

%% file: Sections/simulations.tex

\section{Numerical Simulations}\label{sec:simulation}

\revision{In the previous sections,
	we have presented an estimation-theoretic framework for optimal sensing design
	under resource constraints at processing units and communication channel,
	together with a solution approach based on Reinforcement Learning.
	We next showcase applicability of our setup
	through two edge-computing scenarios.
	This allows us to get insight into the structure of optimal sensing,
	and also shows that our proposed approach can outperform standard design choices}.

\revision{In~\autoref{sec:sim-python}, 
we consider drones for target tracking
and see how online sensor selection can improve performance.
In~\autoref{sec:sim-omnet}, 
we consider smart sensors monitoring an autonomous car
to get insight into processing allocation for heterogeneous networks.}\footnote{
Code available at \url{github.com/lucaballotta/ProcessingNetworks-RL}.}
Finally,
in~\autoref{sec:sim-discussion} we elaborate on the role of Reinforcement Learning
in conjunction with a model-based tool such as Kalman filter. 

%% file: Sections/sim-python.tex

\revision{\subsection{\titlecap{team of drones for target tracking}}}\label{sec:sim-python}

\begin{figure}
	\centering
	\includegraphics[width=0.65\linewidth]{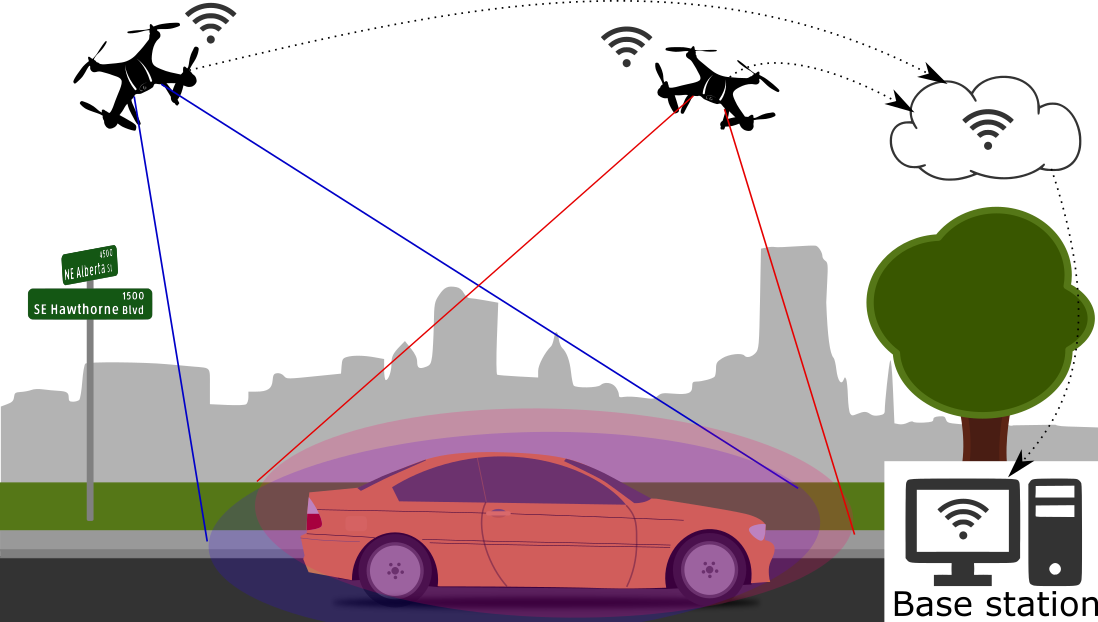}
	\caption{\textbf{Drone tracking simulation setup}.
		The base station estimates the trajectory of the moving target (car)
		based on visual updates from drones.}
	\label{fig:tracking}
\end{figure}

\begin{table}
	\begin{center}
		\small
		\caption{Sensor parameters for drone-tracking scenario. 
		}
		\label{table:params-sensing-2}
		\begin{tabular}{ccccc}
			\toprule
			$ T $ 					 & $ \delayRaw $ 				& $ \delayProc $		 	& $ \varRawS $ 	& $ \varProcS $\\
			$ 10 $\si{\milli\second} &	$ 40 $\si{\milli\second}	& $ 140 $\si{\milli\second}	& $ 10 $	 	& $ 1 $	\\
			\bottomrule
		\end{tabular}
	\end{center}
\end{table}
\begin{table}
	\begin{center}
		\small
		\caption{Learning hyperparameters for drone-tracking scenario.}
		\label{table:params-RL-2}
		\begin{tabular}{cccccc}
			\toprule
			$ M $ & $ \learnRate $ 	& $ \epsmax $ & $ \epsmin $ & $ \epsilon_t $ 			& $ \discFact $\\
			$ 5 $ & $ 0.002 $ 		& $ 0.9 $	  & $ 0.1  $    & 	 {{$\max$} $\left\lbrace\frac{\epsmax}{\sqrt{t+1}},\epsmin \right\rbrace$} 						& $ 1 $ 	   \\
			\bottomrule
		\end{tabular}
	\end{center}
\end{table}
\revision{We simulated}
a team of 25 drones tracking a vehicle on the road (\autoref{fig:tracking})
modeled as a double integrator~\cite{9137405}.
Each drone carries a camera and can either transmit raw frames
or run neural object detection on-board,
sending fairly precise bounding boxes.
We simulated in Python,
with parameters in~\autoref{table:params-sensing-2} based on experiments in~\cite{allan_big_2019,hossain_deep_2019}
and communication delays $ \delayComm = 10\si{\milli\second} $.
We set fusion delays $ \delayFus{k} $ proportional to the number of data that are processed
by Kalman predictor
to compute $ \xhat{k}{} $.
\revision{We addressed an optimization horizon $ [0, K] $ split into ten $ 500\si{\milli\second} $-long windows
	and Q-learning hyperparameters reported in~\autoref{table:params-RL-2},
	where $ t $ means the $ t $th episode (one episode being one horizon),
	training for 500000 episodes.}

\begin{table}
	\revision{\begin{center}
			\small
			\caption{\revision[2]{Network sensing policy $ \policyNet $ learned by Q-learning. 
				The numbers show how many sensors transmit (process) across time.}
			}
			\label{tab:optimal_policies}
			\begin{tabular}{llcc}
				\toprule
				$ v_{\text{proc}} $ &Time window								& 1 		& 2-10 \\
				\midrule
				1					&Transmitting (processing) sensors 			& {10} (0)	& {20} (0)	\\
				\midrule
				0.5 				&Transmitting (processing) sensors 			& {9} (7)	& {11} (11)	\\
				\midrule
				0.1					&Transmitting (processing) sensors 			& {8} (7)	& {14} (14)	\\
				\bottomrule
			\end{tabular}
	\end{center}}
\end{table}

\revision{The sensing design policy learned for the horizon is shown in~\autoref{tab:optimal_policies} (second row).
	Notably,
	only raw mode is chosen:
	in particular,
	10 drones are active in raw mode during the first window
	and 20 through the rest of the horizon.
	This means that, 
	with the parameters in~\autoref{table:params-sensing-2}, 
	data processing at the edge
	is inconvenient because the resource constraints of drones
	induce long processing delays.
	In addition,
	the use of sleep mode,
	that actually implements an \textit{online sensors selection},
	improves performance:
	in words,
	transmission of data from all drones cannot be efficiently handled by 
	the base station and introduces extra computation latency,
	with consequent performance degradation.
	This finding is remarkable because it clashes against the typical assumption
	that performance improves monotonically with the number of sensors}.

\begin{figure}
	\centering
	\includegraphics[width=.8\linewidth]{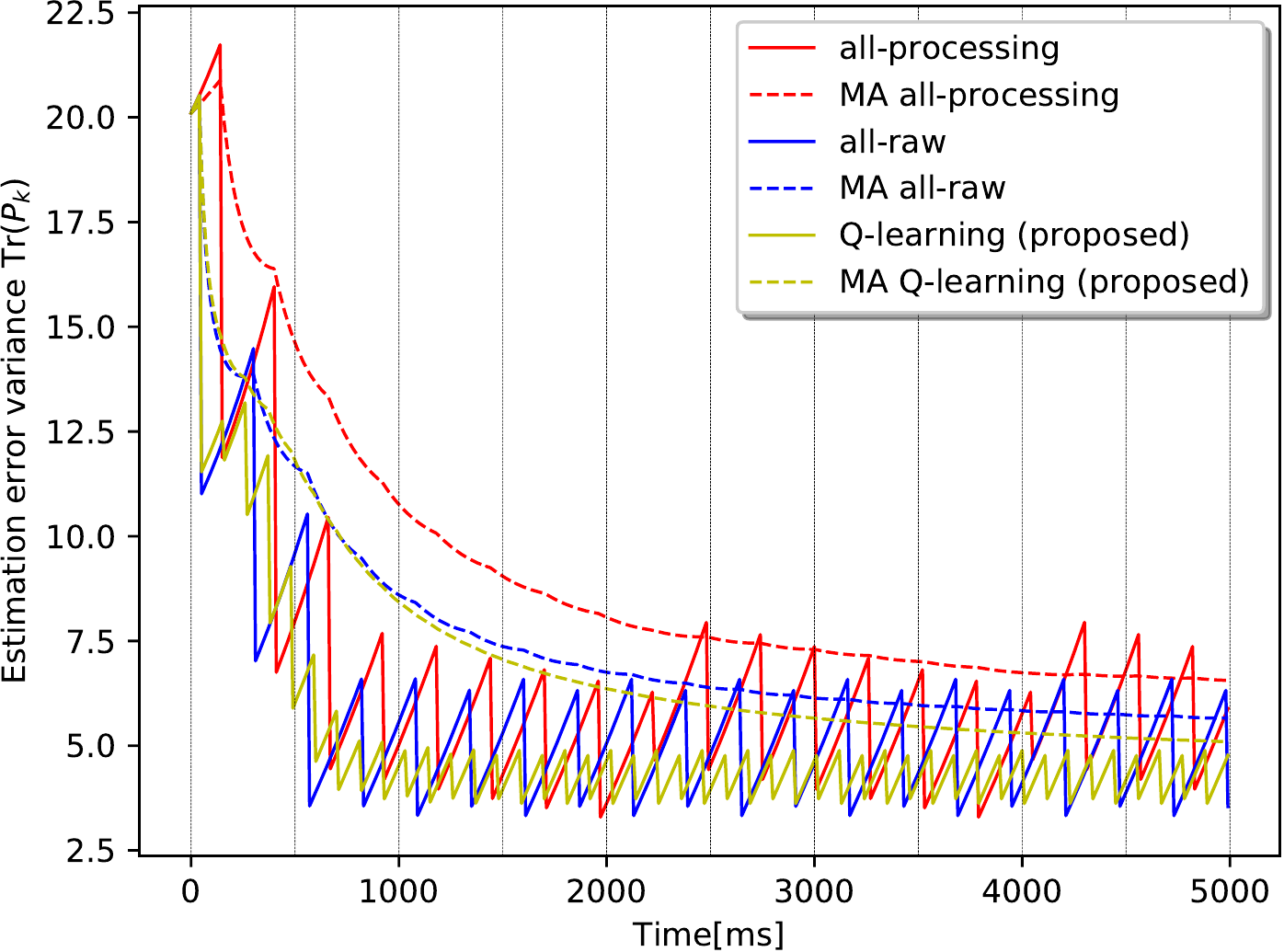}
	\caption{Estimation error variance in drone tracking simulation.}
	\label{fig:comparison_drone_sim}
\end{figure}

\begin{table}
	\begin{center}
		\small
		\caption{Mean error variance in drone-tracking simulation.}
		\label{tab:meanErrVar-2}
		\begin{tabular}{ccc}
			\toprule
			\textbf{Q-learning (proposed)} 	& {All-raw} & {All-processing} \\ 
			\textbf{5.10} 					& {5.69} 	& {6.58} \\
			\bottomrule
		\end{tabular}
	\end{center}
\end{table}

\revision{We compare our approach against two standard, static design choices:
	all sensors transmit raw data at all times (\textit{all-raw})
	and all sensors refine measurements at all times (\textit{all-processing}).}
The comparison 
is shown in~\autoref{fig:comparison_drone_sim} and~\autoref{tab:meanErrVar-2}.
Both baselines are outperformed by optimization~\eqref{eq:prob-optimal-estimation-simplified}.
\revision{Interestingly,
	our solution also keeps small the Moving Average (MA) of the error variance
	with respect to the other two designs}
(see~\autoref{fig:comparison_drone_sim}).
The largest improvement is recorded at steady-state,
while during the transient all curves are very close,
with \textit{all-raw} performing best at times.
This may have two causes:
the transient phase is more difficult to explore for the Q-learning,
but also,
that seemingly sub-optimal behavior during the first two windows might
be necessary given that the learning procedure targets the whole
horizon. 
Indeed, 
the optimal solution to~\eqref{eq:prob-optimal-estimation-simplified}
need not patch together policies that optimize different time windows.

\revision{
	To further investigate the structure of optimal sensing design,
	we have trained policies with different values of sensor parameters.
	In particular,
	we have considered data processing with progressively higher accuracy,
	quantified by measurement noise variances $ \varProcS \in \{1, 0.5, 0.1\} $.
	The learned sensing policies are shown in~\autoref{tab:optimal_policies}.
	As the accuracy of data processing improves,
	fewer sensors are needed to achieve high estimation quality (small error variance),
	while the enhanced processing induces to set more sensors to processing mode.
	This is indeed consistent with intuition and may help in design of real applications.
	Detailed results for the two additional cases are given in Appendix~\ref{app:drone-sim-extra} in Supplementary Material.
}

\begin{figure}
	\centering
	\pgfplotstableread{
		Label xcoord series1 series2
		A 1 498.8 19
		B 1 498.8 0
		C 1.8 378.8 0
	}\testdata
	\begin{tikzpicture}[scale=.7]
		\begin{axis}[
			ybar stacked,
			ymin=0,
			ymax=640,
			xtick=data,
			xticklabels={all-processing, all-raw, Q-learning (proposed)},
			xticklabel style={font = \small},
			ytick={0, 100, 200, 300, 400, 500, 600},
			ylabel=Energy{[\si{\joule}]},
			ylabel style={font = \normalfont},
			ymajorgrids,
			legend style={
				cells={anchor=west},
			},
			reverse legend=true,
			bar width=0.8cm,
			enlarge x limits=0.3
			]
			\addplot [fill=black]
			table [x=xcoord, y=series1, x expr=\coordindex]
			{\testdata};
			\addlegendentry{Sampling and transmission}
			\addplot [fill=black,pattern=north east lines]
			table [x=xcoord, y=series2, x expr=\coordindex]
			{\testdata};
			\addlegendentry{Processing}
		\end{axis}
	\end{tikzpicture}
	\caption{Total energy consumption in drone-tracking simulation.
	}
	\label{fig:energy}
\end{figure}
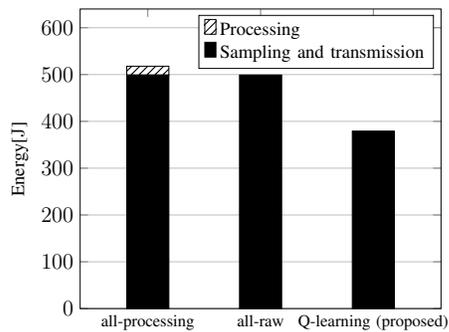

\begin{rem}[Energy \revision{saving}]\label{rem:energy}
	An appealing side effect of \revision{our proposed design through the
	online sensor selection it induces} is reduced energy consumption,
	which can increase the lifespan of the system.
	Considering industrial devices such as Genie Nano cameras~\cite{datasheet_camera},
	with typical power consumption of $ 3.99\si{\watt} $ for sampling and transmission,
	and assuming $ 0.15\si{\watt} $ for data processing~\cite{5306375},
	the energy consumption under the confronted sensing policies is shown in~\autoref{fig:energy} and~\autoref{tab:energyCons}.
	In particular,
	our policy uses 
	only $ 76\% $ of the energy \revision{consumed by \textit{all-raw}}.
\end{rem}

\revision[2]{\begin{rem}[Computational scalability]\label{rem:scalability-qlearning}
	As mentioned in~\autoref{sec:RL-discussion},
	our centralized learning approach can handle small-to-medium network sizes but may struggle when the number of sensors is large.
	To evaluate how the learning complexity scales with the network,
	we have run experiments with various numbers of drones,
	which are reported in Appendix~\ref{app:convergence}.
\end{rem}}

\begin{table}
	\begin{center}
		\small
		\caption{Energy consumption breakdown during transient (trans.,
			\revision{windows 1-2}) and at steady state (ss.,
			\revision{windows 3-10})
			\revision{for sampling and transmission (tx) and processing (proc.)} in drone-tracking simulation.}
		\label{tab:energyCons}
		\begin{tabular}{ccccccc}
			\toprule
			& \multicolumn{2}{c}{\makecell{\textbf{Q-learning}\\\textbf{(proposed)}}} & \multicolumn{2}{c}{{All-raw}} & \multicolumn{2}{c}{{All-processing}}\\
			& \multicolumn{1}{c}{Tx}  & Proc. & {Tx} & Proc. & {Tx} & Proc. \\
			\hline 
			Trans. & \multicolumn{1}{c}{\textbf{59.6\si{\joule}}}	& {0} 	& {99.8}\si{\joule}     & {0} 			& {99.8}\si{\joule} 	& {3.8}\si{\joule} \\
			Ss. & \multicolumn{1}{c}{\textbf{319.2\si{\joule}}}	& {0} 	& {399.0}\si{\joule}    & {0} 			& {399.0}\si{\joule} 	& {15.2}\si{\joule}\\
			\hline
			Total & \multicolumn{2}{c}{{\textbf{378.8\si{\joule}}}} & \multicolumn{2}{c}{498.8\si{\joule}}		& \multicolumn{2}{c}{{517.8\si{\joule}}
			}\\
			\bottomrule
		\end{tabular}
	\end{center}
\end{table}

%% file: Sections/sim-omnet.tex

\subsection{\titlecap{smart sensing for self-driving vehicle}}\label{sec:sim-omnet}

\begin{figure}
	\centering
	\includegraphics[width=0.85\columnwidth]{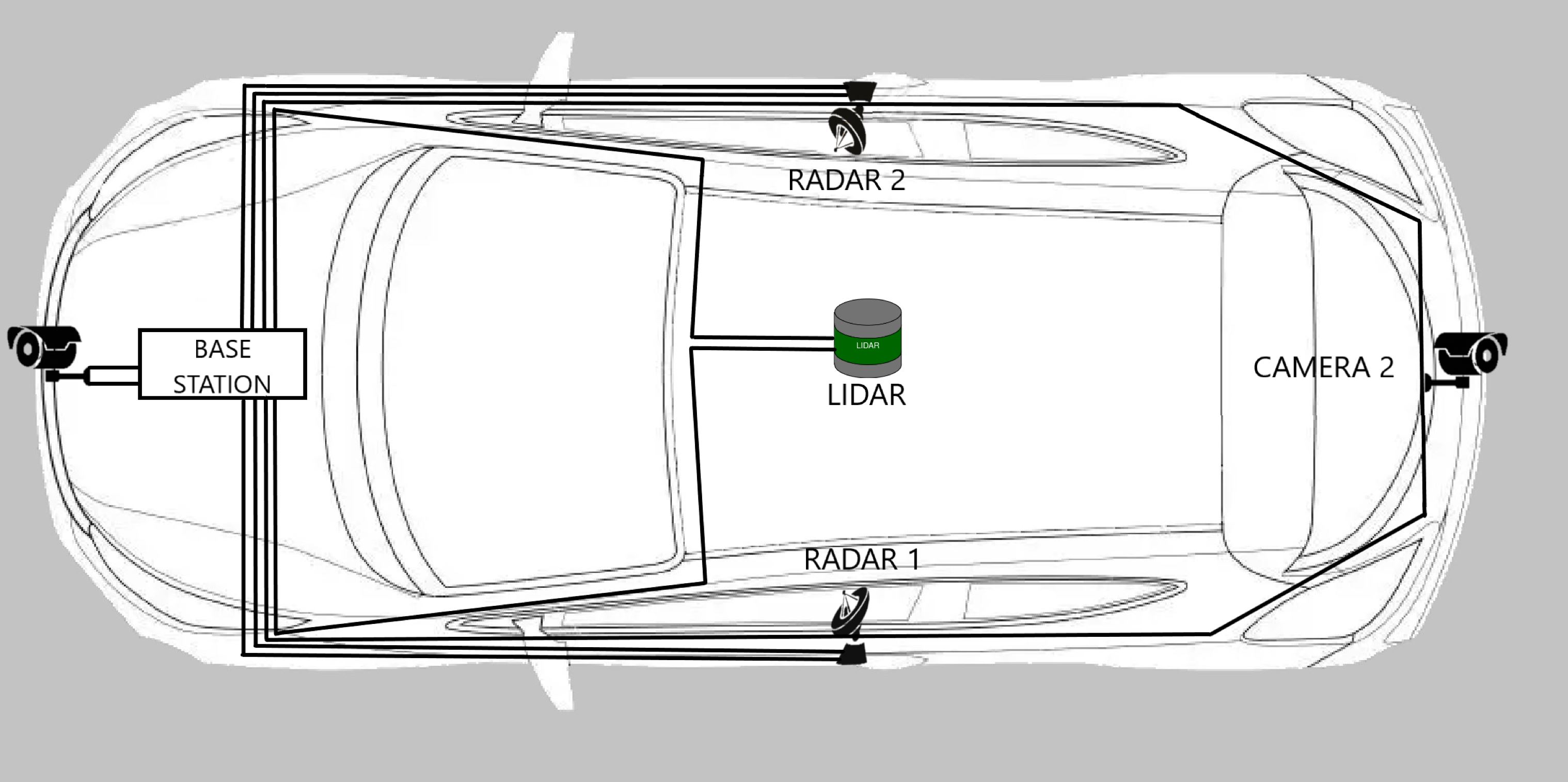}
	\caption{\textbf{Autonomous-driving simulation setup.} 
		Sensors on-board measure the position of the car 
		and a centralized microcontroller tracks its trajectory.}
	\label{simulationNet:fig}
\end{figure}

\revision{In our second experiment,} we considered a self-driving car traveling at approximately constant speed.
Specifically, we considered its transversal position with respect to the center of the lane,
\revision[2]{which is estimated by 
an internal controller (base station)
that receives data from sensors on-board the car and 
tracks the car trajectory
(\autoref{simulationNet:fig}), 
for example to control} a lane shift at sustained speed (\eg for passing or on a highway).
The car dynamics are modeled through a double integrator,
which is a flexible choice used for uncertain dynamics
with direct control of accelerations~\cite{9099596,barreiro_reset_2021,oral_role_2021,rao_naive_2001,ren_consensus_2008}.
Given such a model,
Kalman predictor is an effective estimator
assuming that lateral movements are limited compared to the car speed.

We considered two radar devices, 
two cameras and one lidar,
which are commonly employed in self-driving applications~\cite{9000872}.
Many techniques used in autonomous driving exploit lidar point clouds,
such as segmentation, detection and classification tasks~\cite{9173706}.
Also, 
radars are emerging as a key technology for such systems. 
Some of today’s self-driving cars, \eg Zoox, are equipped with more than 10 
radars providing $360^{\circ}$ surrounding sensing capability under any weather conditions~\cite{9429942}.
Finally, 
camera images are essential to enable 
commercialization of self-driving cars with autonomy at level 3~\cite{9712328}.
The sensor parameters (\autoref{table:params-sensing},
with $  = \varRawS I $ and $ \varProc = \varProcS I $)
were chosen based on real-world experiments~\cite{9127853},
with sampling period $ T=1\si{\milli\second} $ to ensure real-time vehicle control.

\revision{The sensor network is designed according to \revision[2]{the architecture proposed} in~\cite{whitePaperFutureVehicleNetwork}:
	here, 
	smart sensors embed a sensor (\eg camera)
	and a microcontroller
	that can \revision[2]{refine raw} sensory data to decrease the computational effort \revision[2]{for sensor fusion}.
	The \revision[2]{base station} is a controller \revision[2]{inside the car} that manages all jobs \revision[2]{needed for} autonomous driving.
	Because the application is safety-critical,
	transmissions occur through two redundant high-speed Ethernet cables. 
	In light of the small number of sensors and transmission speed,
	we assume that communication latency is negligible with respect to sampling and processing.}

\begin{table}
	\begin{center}
	\small
		\caption{Sensor parameters for autonomous driving scenario. 
		}
		\label{table:params-sensing}
		\begin{tabular}{cccccc}
			\toprule
			&  Freq.	   	& $ \delayRaw $ 				& $ \delayProc $		 	& $ \varRawS $ & $ \varProcS $\\
			Radar	& $ 50 $\si{\hertz} & 		$ 20$\si{\milli\second}					& $ 30 $\si{\milli\second}	& 	$0.45$	 	   & 	$0.40$		\vspace{0.1cm}	 \\
			Camera	& $ 25 $\si{\hertz} & $ 40 $\si{\milli\second}		& $ 100 $\si{\milli\second}	& 		$0.30$		 	   & 	$0.05$		\vspace{0.1cm}	 \\ 
			Lidar	& $ 10 $\si{\hertz}	& $ 100 $\si{\milli\second}		& $ 140 $\si{\milli\second}	& 		$0.10$		 	   & 	$0.05$		\vspace{0.1cm}	 \\			
			\bottomrule
		\end{tabular}
	\end{center}
\end{table}

\begin{table}
	\begin{center}
    	\small
    	\caption{Learning hyperparameters for autonomous-driving scenario.}
    	\label{table:params-RL-1}
    	\begin{tabular}{cccccc}
    		\toprule
    		$ M $ & $ \learnRate $ 	& $ \epsmax $ & $ \epsmin $ & $ \epsilon_t $ 			& $ \discFact $\\
    		$ 5 $ & $0.1  $ 		& $ 0.2 $	  & $0.01  $     & 	 {{$\max$} $\left\lbrace\frac{\epsmax}{\sqrt{t+1}},\epsmin \right\rbrace$} 						& $ 1 $ 	   \\
    		\bottomrule
    	\end{tabular}
    \end{center}
\end{table}
Communication was simulated through the discrete-event simulator Objective Modular Network Testbed in C++ (OMNeT++) \cite{OMNeT}. 
This is widely adopted to simulate 
networks,
because it combines standard communication protocols (\eg\xspace\revision{IEEE 802.3})
and the possibility to create customized procedures exploiting existing modules.
Further,
it enables realistic simulations by accurately modeling 
both the electromagnetic environment and the lower layers of the protocol stack (from physical to transportation layers).
In our simulations,
sensors carry \revision{IEEE 802.3 (so-called Ethernet)} communication boards.

\revision{For training,
we considered a time horizon $ [0, K] $
split into five time windows
with length $ 300 $\si{\milli\second} each,
and trained for
$ 100000 $ episodes
with hyperparameters reported in~\cref{table:params-RL-1}.}

\begin{table}
	\centering
	\caption{Q-learning policy for autonomous-driving scenario.}
	\includegraphics[width=.8\linewidth]{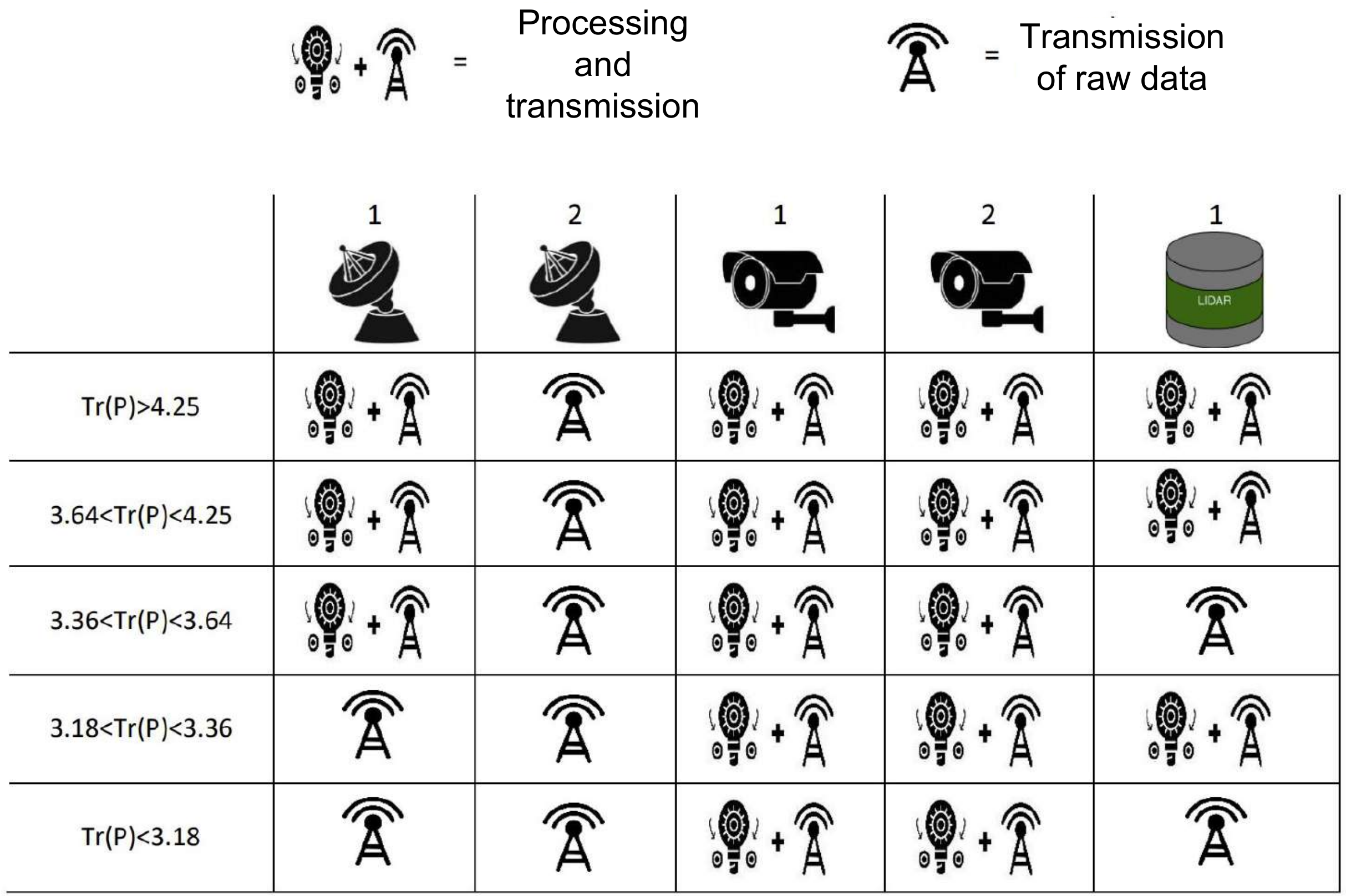}
	\label{tab:tabPolicy}
\end{table}

From ~\autoref{tab:tabPolicy} we can infer that the learned policy requires processing
almost from all the sensors when the error variance is high (top row). 
However,
the need for processing diminishes with the variance, 
turning to raw mode both lidar and radars at the smallest values (bottom row).
Interestingly, processing mode is always chosen for cameras,
revealing that refining of image frames overhangs the additional computational delay. 
Note that in this case,
given the small amount of sensors,
the fusion delays induced at the base station are negligible
and sleep mode is never selected,
namely, sensors always transmit.

\begin{figure}
	\centering
	\includegraphics[width=0.8\columnwidth]{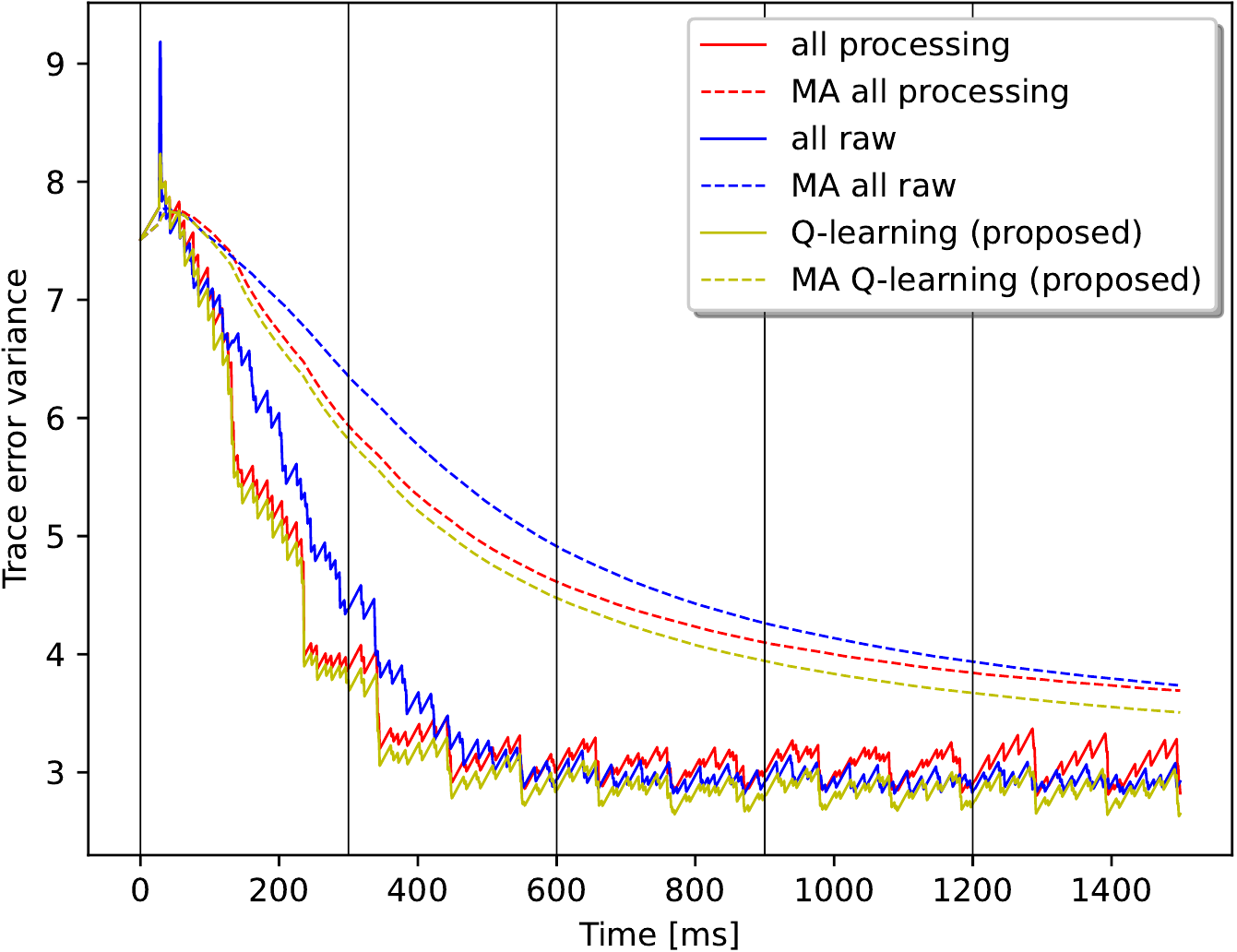}
	\caption{Estimation error variance in autonomous-driving simulation.}
	\label{fig:result2}
\end{figure}

\begin{table}
	\begin{center}
	    \small
		\caption{Mean error variance in autonomous-driving simulation.}
		\label{tab:meanErrVar}
		\begin{tabular}{ccc}
			\toprule
			\textbf{Q-learning (proposed)} 	& {All-raw} & {All-processing} \\ 
			{\textbf{\revision{3.51}}} 				& \revision{3.73}	& \revision{3.69} \\
			\bottomrule
		\end{tabular}
	\end{center}
\end{table}

\revision{The learned policy was tested 
	against the two standard design choices 
	\textit{all-raw}
	and \textit{all-processing} like the previous scenario.
	The outcome over the horizon is plotted in~\autoref{fig:result2}
	and summarized in~\autoref{tab:meanErrVar}.
	As it is possible to appreciate from~\autoref{fig:result2},
	the Q-learning learns to cleverly allocate computational resources
	according to the current estimate accuracy.}
During the transient phase (till $ 600\si{\milli\second} $),
when the error variance is large,
processing mode is selected for lidar, 
cameras and one radar, according to the first two rows in~\autoref{tab:tabPolicy}.
Notably,
this choice performs close to \textit{all-processing} (red curve),
while the \textit{all-raw} configuration is clearly disadvantageous (higher blue curve).
Conversely,
at steady state 
only the cameras are in processing mode:
this resembles more closely the \textit{all-raw} policy,
which performs better (lower blue curve) than \textit{all-processing}.

Overall,
we can see from~\cref{tab:meanErrVar} that the proposed approach leads to a total improvement of about $ 5\% $
compared to baseline policies.
While this result may look marginal,
we note that the improvement is rather small over the main transient phase,
because the Kalman predictor is able to drop the error variance very quickly for all sensing configurations,
but is way larger (about $ 15-20\% $) when the curves settle about small values.
Also,
while the
objective cost~\eqref{eq:prob-optimal-estimation-simplified-objective} refers to the whole horizon,
we note that in fact the learned policy performs better than the baselines nearly at each point in time,
as~\autoref{fig:result2} shows,
with the curve obtained with the Q-learning policy being almost always below the others.
Further,
the MA is again consistently smaller than both baselines,
highlighting an even better performance of the proposed approach with respect to the targeted optimization.

%% file: Sections/sim-discussion.tex

\subsection{\titlecap{discussion: the role of learning in model-based estimation}}\label{sec:sim-discussion}

The \revision{exposed} simulations \revision{suggest} that the \revision{proposed} approach can improve
performance of smart sensor networks dealing with estimation tasks
\revision{as compared to standard design choices with static processing decisions.}
\revision{In particular,
	the learning-based design
	exploits observation of the estimation error online 
	to select effective sensing configurations at different points in time, 
	while baselines cannot adapt to transient or steady-state regimes
	that benefit from different processing allocations.}

It is noteworthy that a learning method such as Q-learning
can effectively drive the sensing design,
leading to improvement with respect to baselines,
even with an estimation tool as effective and robust as the Kalman predictor.
Indeed, \revision{due to optimality of the latter algorithm applied to the chosen dynamical system, one can expect} 
even trivial choices (such as \textit{all-raw} and \textit{all-processing}) to yield acceptable performance. 
\revision{Conversely, it is hard to suggest good heuristics in the present framework,
as the performance varies with respect to the system dynamics, delays, and error variances.}
\revision{In particular,}
an \textit{optimal} design given all available options is far from trivial:
even the simplest setup bears a combinatorial problem that quickly makes
deriving an optimal solution computationally infeasible.
Indeed, submodularity properties
that allow to analytically bound suboptimality of greedy algorithms~\cite{9099596}
are hard to meet in realistic scenarios,
\eg under delays,
out-of-sequence message arrivals,
or multi-rate sensors~\cite{9137405}.

Given these premises,
the performance improvements obtained via the studied learning method are encouraging not only with regard to the \revision{addressed framework},
but mostly in supporting the contribution of such tools
to general estimation and control tasks,
which can benefit from the power of learning to circumvent computational bottlenecks associated with optimization-based design.
Hence,
rather than looking at the two domains of model-based and data-driven control as mutually exclusive approaches,
this work aims to reinforce arguments supporting a unified,
best-of-both-worlds framework.

%% file: Sections/conclusion.tex

\section{Conclusion}\label{sec:conclusion}

Motivated by smart sensing for Edge Computing, 
we have proposed an adaptive design that addresses impact
of resource-constrained data sampling,
processing,
and transmission on performance of a monitoring task.
Starting from a suitable mathematical model for the considered class of systems,
we have tackled the sensing design problem via Q-learning, 
showing that the learned design can considerably improve performance
compared to standard configurations \revision{that do not adapt to the time evolution of the system}.

Future research avenues are multifold.
\revision{Besides challenges of Reinforcement Learning (see~\autoref{sec:RL-discussion}),
model assumptions may be adjusted
to address more realistic sensing and communication,
as well as different dynamics or control tasks.
Also,
our approach should be validated with real-world data.}

%% file: processing_networks_estimation_RL.bbl

%% file: Bio/bios.tex

\begin{IEEEbiography}[{\includegraphics[width=1in,height=1.25in,clip,keepaspectratio]{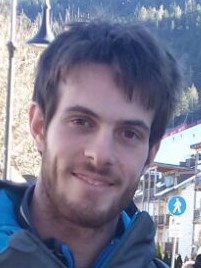}}]{Luca Ballotta}
	\input{Bio/ballotta.txt}
\end{IEEEbiography}

\begin{IEEEbiography}[{\includegraphics[width=1.2in,height=1.75in,clip,angle=270,keepaspectratio]{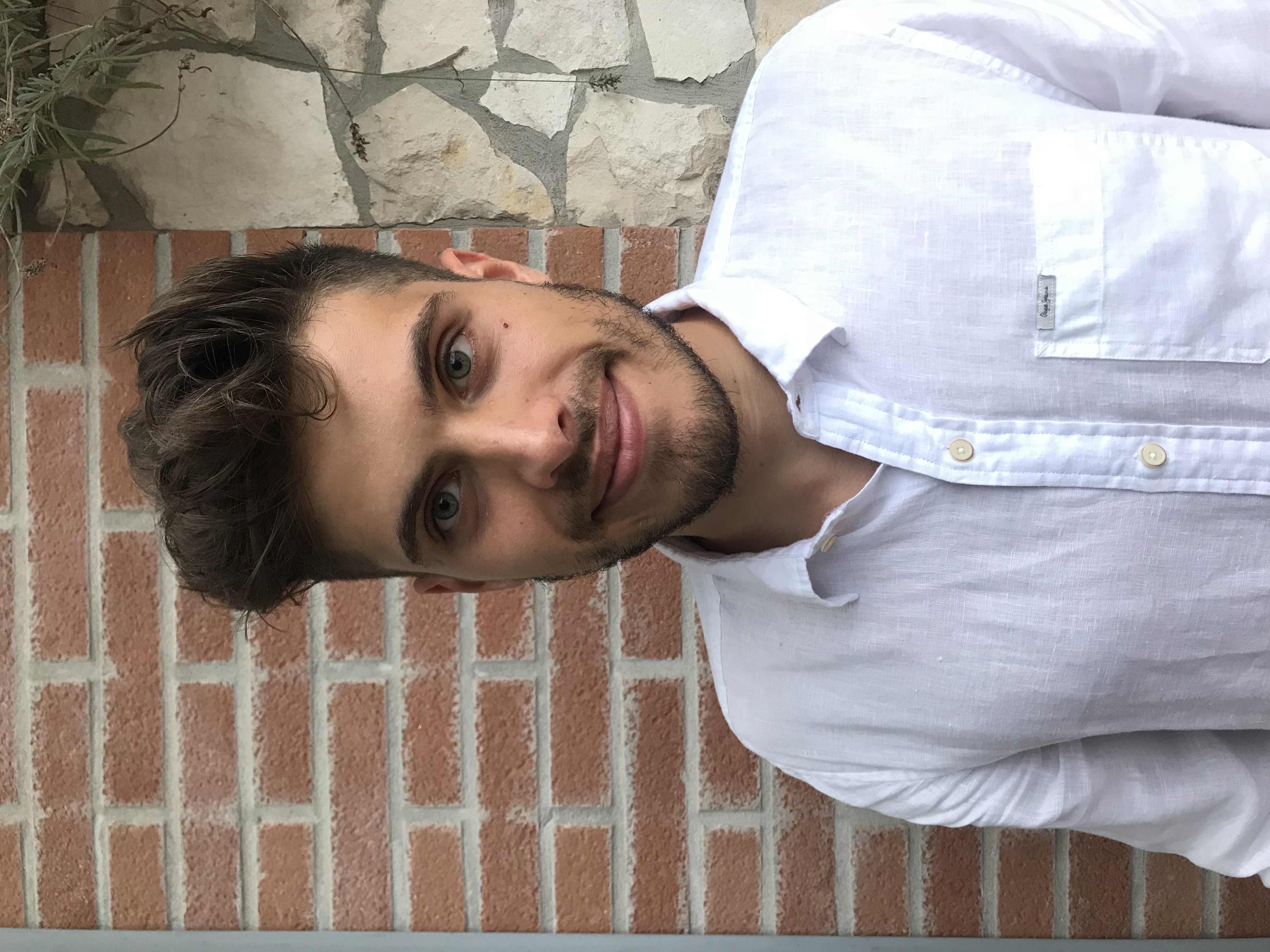}}]{Giovanni Peserico}
	\input{Bio/peserico.txt}
\end{IEEEbiography}

\begin{IEEEbiography}[{\includegraphics[width=1in,height=1.25in,clip,keepaspectratio]{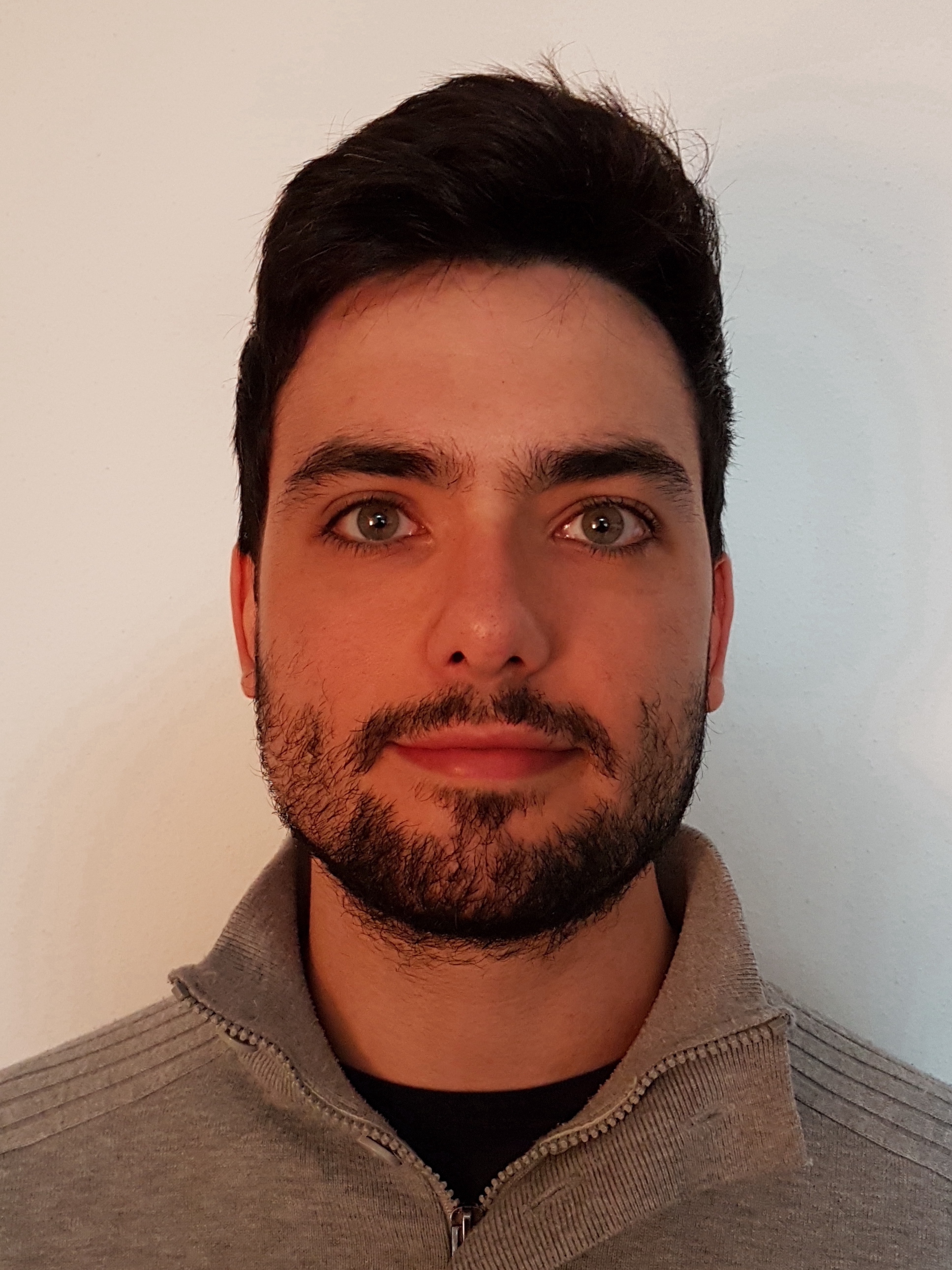}}]{Francesco Zanini}
	\input{Bio/zanini.txt}
\end{IEEEbiography}

\begin{IEEEbiography}[{\includegraphics[width=1in,height=1.25in,clip,keepaspectratio]{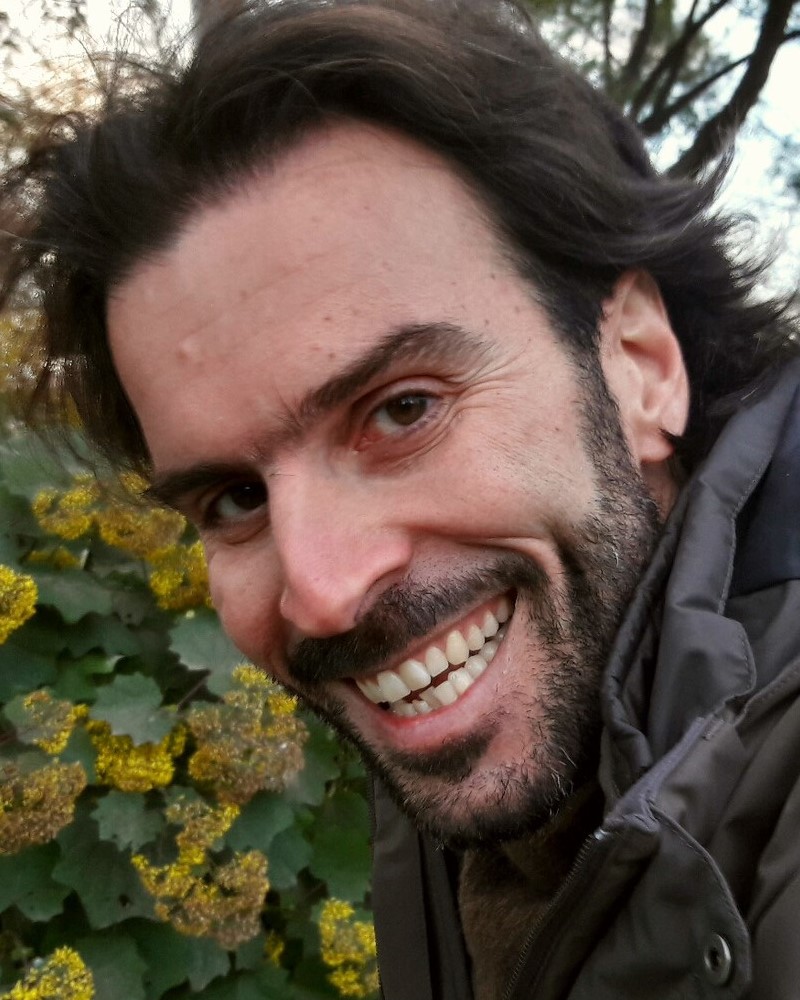}}]{Paolo Dini}
	\input{Bio/dini.txt}
\end{IEEEbiography}

%% file: Bio/ballotta.TXT
received the Master's degree in Automation Engineering and the Ph.D. degree in Information Engineering from the University of Padova, Italy, in 2019 and 2023, respectively.
He was Visiting Student at the Massachusetts Institute of Technology in 2020 and 2022.
He was awarded with the Young Author Prize at the 2020 IFAC World Congress.
His research interests include multi-agent systems and networked control systems subject to resource constraints, 
resilient distributed optimization,
and learning-based safe control.

%% file: Bio/peserico.TXT
received the Master's degree in Automation Engineering and the Ph.D. degree in ``Alto Apprendistato'' in Information Engineering in collaboration with Autec s.r.l. from the University of Padova, Italy, in 2019 and 2023, respectively.
He is now covering a Cybersecurity Software Engineer position for Qascom, an italian company specialized in GNSS authentication and space cybersecurity.
His research interests include safety and cybersecurity, industrial and wireless networks,  networked control systems
and learning-based safe control.

%% file: Bio/zanini.TXT
received the Master's degree in Automation Engineering and the Ph.D. degree in Information Engineering from the University of Padova, Italy, in 2019 and 2023, respectively.
He was Visiting Student at the University of Alberta in 2022, and later joined the institution as a post-doctoral researcher in 2023. 
His research interests lie at the intersection of reinforcement learning and dynamical systems, along with Koopman operators and learning theory.

%% file: Bio/dini.TXT
received the M.Sc. and Ph.D. degrees from the Universit\`a di Roma La Sapienza in 2001 and 2005, respectively. He is currently a Senior Researcher with the Centre Tecnol\'ogic de Telecomunicacions de Catalunya (CTTC), where he coordinates the activities of the Sustainable Artificial Intelligence research unit. His research interests include sustainable computing and networking, distributed optimization and machine learning, multi-agent systems and decision-making processes, data mining for cyber-physical systems. He has been involved in more than 25 research projects during his career. He is currently the Coordinator of the CHIST-ERA SONATA project on sustainable computing and communication at the edge and the Scientific Coordinator of the MSCA Greenedge European Training Network on edge intelligence and sustainable computing. His research activity is documented in more than 90 peer-reviewed scientific journals and international conference papers. He received two awards from the Cisco Silicon Valley Foundation for his research on heterogeneous mobile networks in 2008 and 2011, respectively. He has co-organized several training events and workshops/special sessions at several international conferences sponsored by IEEE. He serves as a TPC in many international conferences and a reviewer for several scientific journals of the IEEE, Elsevier, ACM, Springer. He is European Climate Pact Ambassador since 2022 and participates in several outreach events (e.g., Research Nights) to promote sustainable design principles.

%% file: Appendix/kalman-filter.tex

\section{\titlecap{kalman predictor with delayed updates}}\label{app:kalman-filter}

Assume that at time $ k - \delayFus{k} $ the base station can use the following
time-sorted measurements to compute $ \xhat{k}{} $, 
\begin{equation}\label{eq:all-measurements-sorted}
	\allmeasurements{k} = \lb\lr y_{k_0},V_{k_0}\rr,\dots,\lr y_{k_M},V_{k_M}\rr \rb, \quad k_i < k_{i+1},
\end{equation}
with $ k_M \le k -\delayFus{k} $.
The following procedure can handle out-of-sequence measurements sampled at or after time $ k_0 $ (oldest sample in~\eqref{eq:all-measurements-sorted})
and received before or at time $ k - \delayFus{k} $.
For the sake of clarity, in~\eqref{eq:all-measurements-sorted} we have omitted subscripts and superscripts
related to sensors.
The estimation error covariance associated with $ \xhat{k}{} $
given by Kalman predictor starting from $ P_{k_0} $ and using measurements in $ \allmeasurements{k} $ is given by~\cite{9137405}
\begin{equation}\label{eq:Kalman-filter}
	\Pmat{k}{} = \pred{\update{...\update{\pred{\update{P_{k_0}}{V_{k_0}}}{k_0:k_1}}{V_{k_1}}{...}}{V_{k_M}}}{k_M:k},
\end{equation}
where the multi-step open-loop update between time $ k_i $ and time $ k_j \ge k_i $
(due to lack of measurements in $ (k_i,k_j) $) is
\begin{equation}\label{eq:Kalman-filter-prediction-update}
	\begin{aligned}
		&\pred{P}{k_i:k_j} = \mathcal{P}_{k_j}\circ\dots\circ\mathcal{P}_{k_i}(P), 
		\quad \pred{P}{k_i:k_i} \doteq P\\
		&\mathcal{P}_{k_i}\left(P\right) \doteq A_{k_i}PA_{k_i}^\top + W_{k_i},
	\end{aligned}
\end{equation}
and the update with the $ i $th measurement sampled at time $ k_i $ is
\begin{equation}\label{eq:Kalman-filter-update}
	\update{\Pmat{k_i}{}}{\V{k_i}{}} = \left((\Pmat{k_i}{})\inv + \left(\V{k_i}{}\right)\inv\right)\inv.
\end{equation}

%% file: Appendix/drone-sim-extra.tex

\begin{figure}[b]
	\centering
	\includegraphics[width=.8\linewidth]{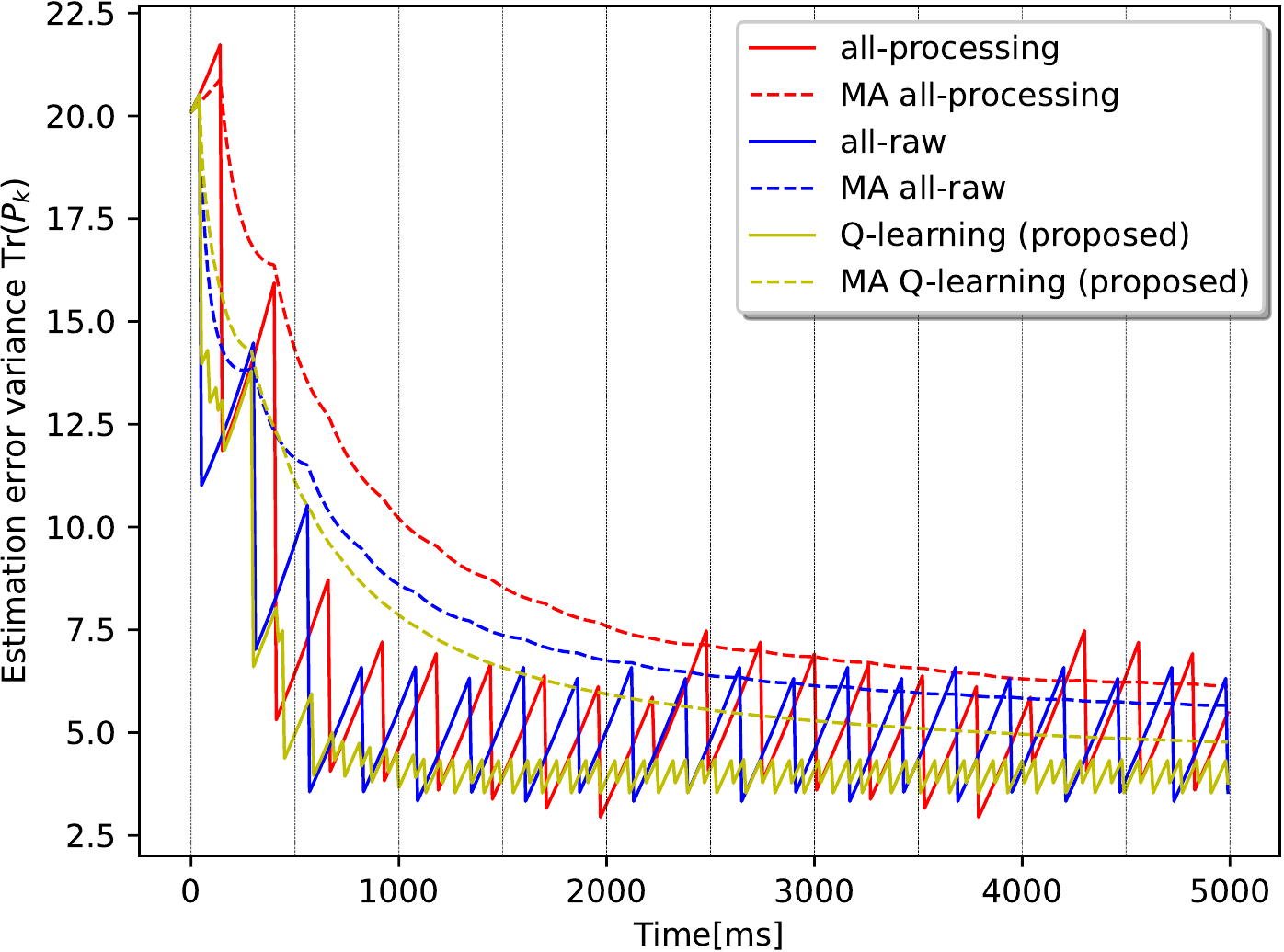}
	\caption{\revision{Error variance in drone-based tracking simulation with $ v_\text{proc} = 0.5 $.}}
	\label{fig:comparison_drone_sim_0.5}
\end{figure}

\begin{figure}
	\centering
	\includegraphics[width=.8\linewidth]{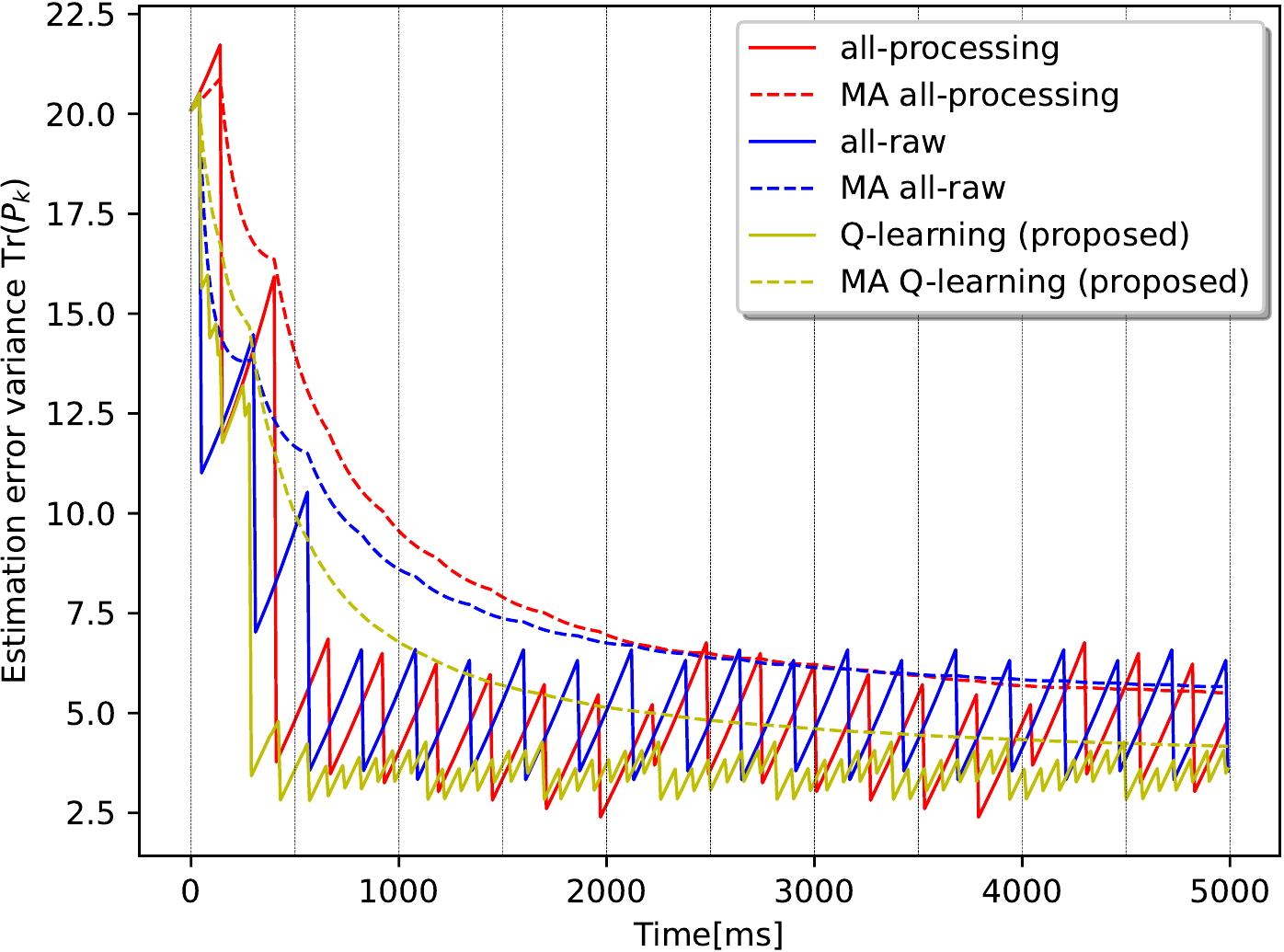}
	\caption{\revision{Error variance in drone-based tracking simulation with $ v_\text{proc} = 0.1 $.}}
	\label{fig:comparison_drone_sim_0.1}
\end{figure}

\revision{\section{Team of Drones for Target Tracking: Additional Simulation Results}\label{app:drone-sim-extra}

In this appendix,
we report additional experiments where we investigate how the sensing design by Q-learning varies
as we change the measurement noise variance of processed data,
\ie the accuracy of data provided by sensors in processing mode.

\autoref{tab:meanErrVar-all} summarizes performances of the optimized sensing policies
against the two baselines \textit{all-raw} and \textit{all-processing}.

Figures~\ref{fig:comparison_drone_sim_0.5} and~\ref{fig:comparison_drone_sim_0.1}
show the behavior of the error variance along the addressed horizon $ [0,K] $
when the measurement noise of processed data has variance $ \varProcS $ equal to $ 0.5 $ and $ 0.1 $,
respectively.

Finally,
\autoref{fig:energy-all} shows the estimated energy consumption under the considered sensing designs,
which is significantly reduced by the adaptive policies learned through our approach.
This demonstrates an attractive by-product of a careful design
that takes into account model features such as latency and accuracy of supplied sensory data.
\autoref{tab:energyCons-all} reports in detail the energy consumption under the learned sensing policies
for the different cases of $ \varProcS $
over transient and steady-state.}

\begin{table}[b]
	\revision{\begin{center}
		\small
		\caption{Mean error variance in drone-tracking simulation.
			Figures between brackets show cost decrease w.r.t. second best policy.
		}
		\label{tab:meanErrVar-all}
		\begin{tabular}{cccc}
			\toprule
			$ v_\text{proc} $		& Q-learning (proposed)				 	& {All-raw} & {All-processing} \\ 
			\midrule
			1 						& \textbf{5.10} (-10\%)					& {5.69} 	& {6.58} \\
			0.5 					& \textbf{4.77} (-16\%)					& {5.69} 	& {6.16} \\
			0.1 					& \textbf{4.17} (-24\%)					& {5.69} 	& {5.53} \\
			\bottomrule
		\end{tabular}
	\end{center}}
\end{table}

\begin{figure}
	\centering
	\pgfplotstableread{
		Label	xcoord series1	series2
		A		1		498.8		19
		B		2		498.8		0
		C		3		378.8		0
		D 		4		215.5		8.1
		E 		5		267.3		10.0
	}\testdata
	\revision{\begin{tikzpicture}[scale=0.8]
		\begin{axis}[
			ybar stacked,
			ymin=50,
			ymax=540,
			xmin=0,
			xmax=4,
			xtick=data,
			xticklabels={all-processing, all-raw, \makecell{Q-learning \\ ($ v_\text{proc} = 1 $)},
				\makecell{Q-learning \\ ($ v_\text{proc} = 0.5 $)}, \makecell{Q-learning \\ ($ v_\text{proc} = 0.1 $)}},
			xticklabel style={font = \footnotesize},
			ytick={100, 200, 300, 400, 500, 600},
			ylabel=Energy{[\si{\joule}]},
			ylabel style={font = \small},
			ymajorgrids,
			legend style={font = \small,
				cells={anchor=west},
			},
			reverse legend=true,
			bar width=0.5cm,
			enlarge x limits=0.1,
			width=9cm
			]
			\addplot [fill=black]
			table [x=xcoord, y=series1, x expr=\coordindex]
			{\testdata};
			\addlegendentry{Sampling and transmission}
			\addplot [fill=black,pattern=north east lines]
			table [x=xcoord, y=series2, x expr=\coordindex]
			{\testdata};
			\addlegendentry{Processing}
		\end{axis}
	\end{tikzpicture}}
	\caption{\revision{Total energy consumption in drone-tracking simulation.}}
	\label{fig:energy-all}
\end{figure}



\begin{table}
	\revision{\begin{center}
		\small
		\caption{Energy consumption breakdown during transient (trans.,
			windows 1-2) and at steady state (ss.,
			windows 3-10)
			for sampling and transmission (tx) and processing (proc.) in drone-tracking simulation.}
		\label{tab:energyCons-all}
		\begin{tabular}{ccccccc}
			\toprule
			\multirow{2}{*}{\makecell{Q-learning \\ (proposed)}}	& \multicolumn{2}{c}{$ v_\text{proc} = 1 $} & 
			\multicolumn{2}{c}{$ v_\text{proc} = 0.5 $} 			& \multicolumn{2}{c}{$ v_\text{proc} = 0.1 $} \\
					& Tx	& Proc.			& Tx 				& Proc. 		& Tx 				& Proc. \\
			\hline 
			Trans. 	& 59.6\si{\joule} 			& 0				& 39.9\si{\joule}	& 1.5\si{\joule}& 43.9\si{\joule}	& 1.6\si{\joule}	\\
			Ss. 	& 319.2\si{\joule} 			& 0				& 175.6\si{\joule}	& 6.6\si{\joule}& 223.4\si{\joule}	& 8.6\si{\joule}	\\
			\hline
			Total	& \multicolumn{2}{c}{378.8\si{\joule}} & \multicolumn{2}{c}{223.6\si{\joule}} & \multicolumn{2}{c}{277.3\si{\joule}}	\\
			\bottomrule
		\end{tabular}
	\end{center}}
\end{table}

%% file: Appendix/Qconvergence.tex
\revision[2]{
	\section{\titlecap{heuristic convergence of q-learning}}\label{app:convergence}
	
	Although convergence of the Q-learning algorithm has been theoretically established in \cite{NIPS1993_5807a685}, this result holds for Markov Decision Processes, while the setting under investigation in this paper does not enjoy the Markov property because of the entanglement between the estimator's dynamics and the sensors' operations. For instance, when the transition between two windows happens, the current measurements which still have to be delivered are discarded (see~\autoref{fig:delayed-meas-centr}): the state alone cannot account for this process as it is defined, so the dynamics considered in the formulation of Reinforcement Learning is not Markovian. One may then wonder if the Q-learning approach is flexible enough for this problem, and in particular if it converges around a fixed policy within the number of episodes that have been considered. Due to the non-Markovianity of the environment the answer can only be provided through numerical experiments.
	
	In this appendix,
	we show the convergence results for several training instances related to the first setting (homogeneous network),
	in which a team of drones is tracking a vehicle (see~\autoref{sec:sim-python} and~\cref{app:drone-sim-extra}).
	
	\begin{figure}
		\centering
		\includegraphics[width=.8\linewidth]{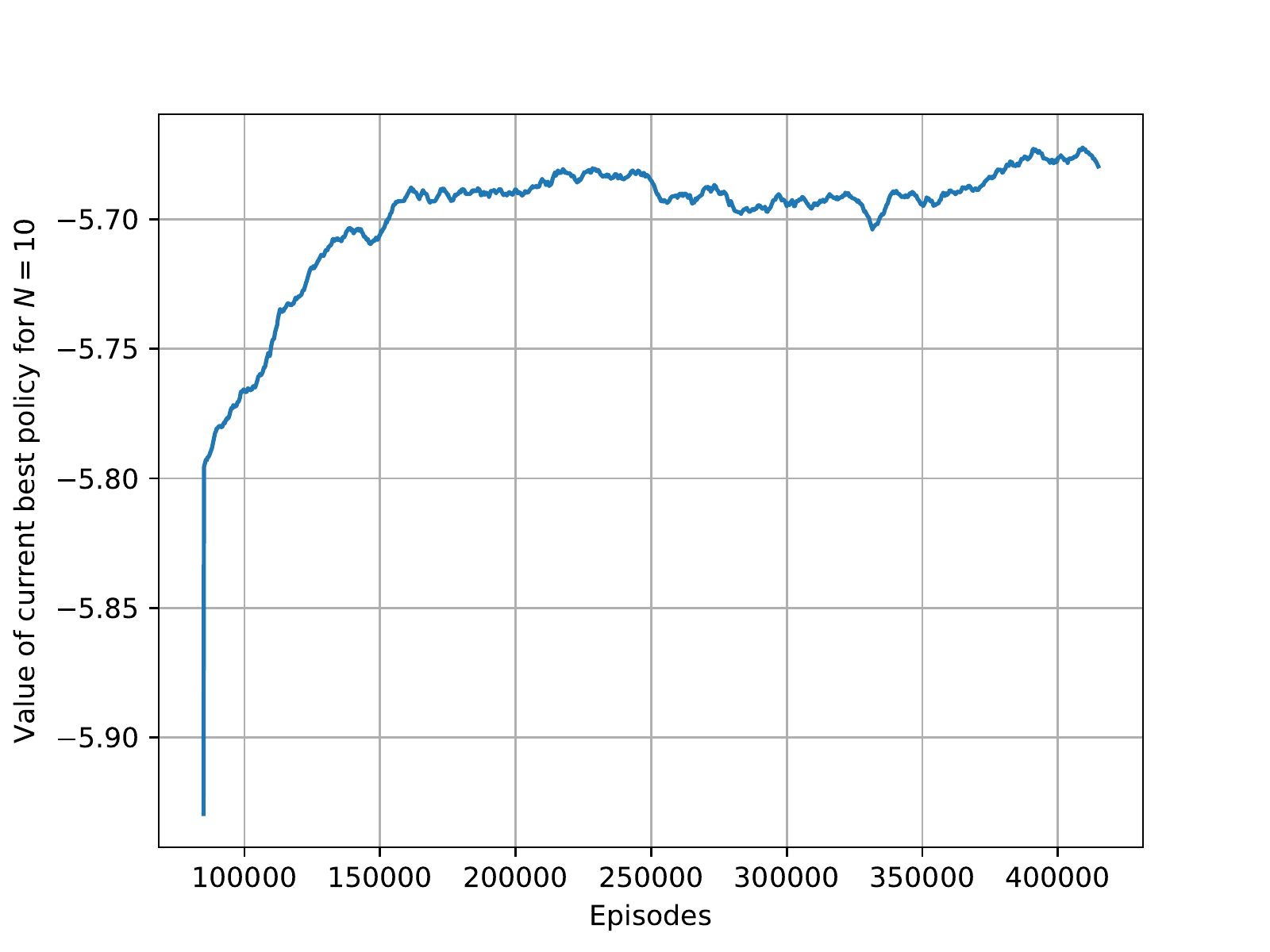}
		\caption{\revision[2]{Values of the optimal policy under the current Q-table during training, for $N=10$}.}
		\label{fig:conv10}
	\end{figure}
	
	\Cref{fig:conv10} shows the behaviour of the long-term reward (the negative sum of the traces of the error covariance over the different windows within an episode -- see Problem~\eqref{eq:RL-cost-specialized}) of the \emph{supposedly} optimal policy under the \emph{current} Q-table. By interacting with the environment and trying different sensors configurations the algorithm is learning a more reliable Q-table, whose maximisation leads to a policy which performs better and better on the true environment, as it is clear from the plot. For the simple case of $N=10$ sensors, the algorithm reaches an empirically stable value already around $175000$ episodes. 
	
	In order to understand how the sample complexity of the algorithm scales with the number of sensor, the same graph is drawn for the case of $N=25$ and $N=35$ sensors in~\autoref{fig:conv-comp}.
	
	\begin{figure}
		\centering
		\includegraphics[width=.8\linewidth]{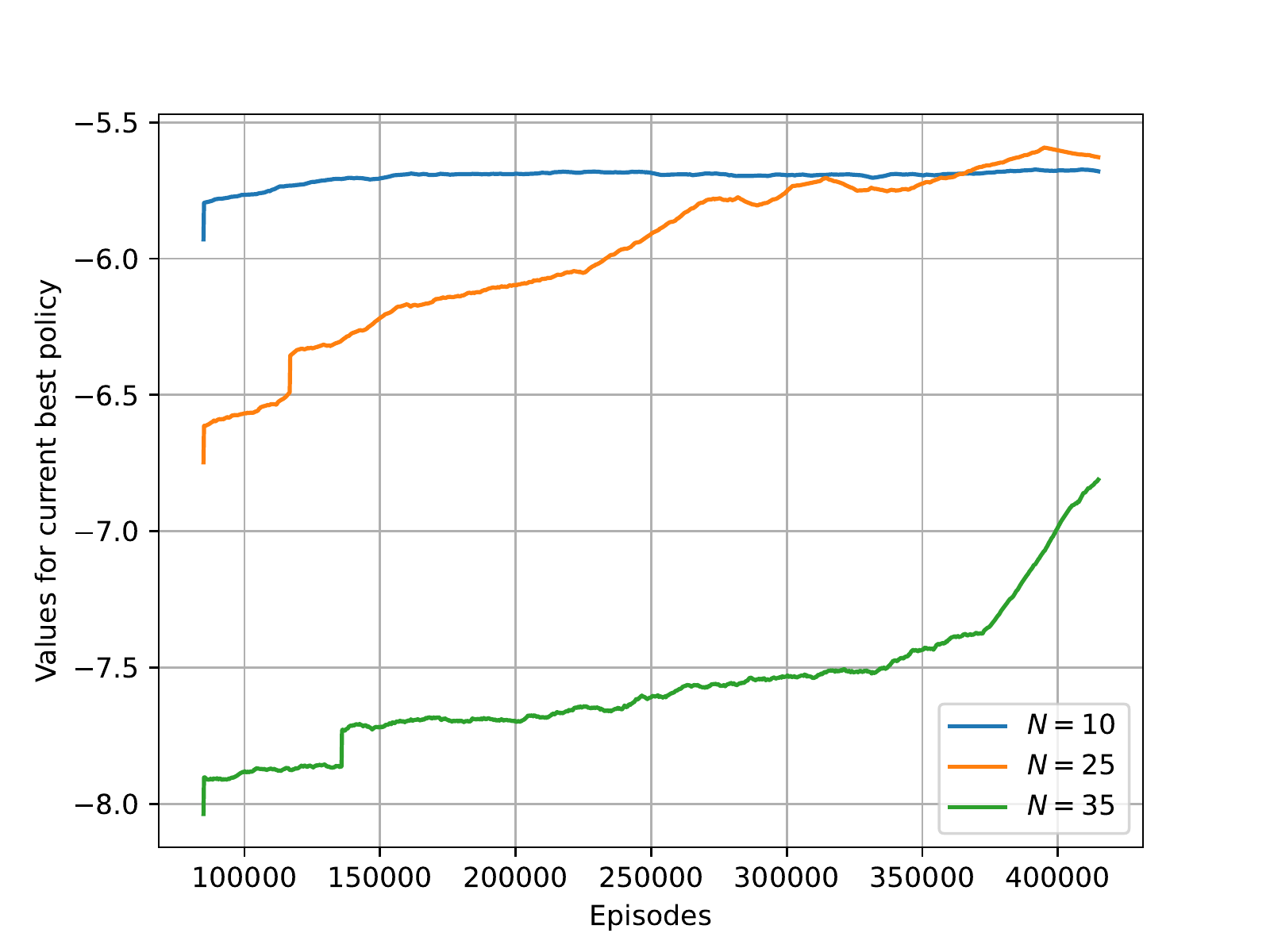}
		\caption{\revision[2]{Comparison of the training performances among different number of sensor available, $N\in\left\{ 10, 25, 35\right\}$}.}
		\label{fig:conv-comp}
	\end{figure} 
	
	The curve with $N=25$ starts to converge around episode $400000$, while the one with $N=35$ 
	does not reach a visible convergence within the chosen training horizon ($500000$ total episodes),
	corresponding to a superlinear trend. 
	This can be understood from the fact that the size of the action set scales quadratically with the number of sensors (as there are $3$ configurations for each sensor) and the latest proposed bound for the sample complexity of Q-learning \cite{pmlr-v139-li21b} establishes a linear dependence in the size of the action set, therefore leading to a quadratic dependence on the number of sensors in our scenario. 
	Note however that by considering a higher number of sensor the optimal policy necessarily attains a steady-state value which is equal or higher to the one which considers fewer sensor: adding more sensor cannot tamper with performance since,
	in the worst case,
	these additional sensors can be put in sleep mode. 
	Therefore,
	the green curve will eventually reach the same value as the orange one, or higher. 
	
	In the case of heterogeneous sensors, the size of the action set scales exponentially with the number of different kind of sensors available, thus making it computationally very expensive to perform a numerical analysis on the convergence of the algorithm. Note also that such an analysis would need to take care of both the total number of sensors and the number of different kind of sensors available, involving many more experiments than the homogeneous case. An extensive convergence analysis is out of the scope of the present contribution. 
	
	Recall also that the case of many different (heterogeneous) sensors is quite atypical in real-world large-scale control applications, 
	which usually comprise only a few types of sensors,
	which therefore makes the computational scalability -- and the related convergence analysis -- similar to the homogeneous case.
}